\documentclass{elsart}

\usepackage{psfig,latexsym,amssymb}

\newcommand{\figwidth}{0.9\columnwidth}

\begin{document}

\begin{frontmatter}

  \title{A supersymmetric
    \protect\boldmath{$\mbox{U}_{q}[\mbox{osp}(2|2)]$}-extended \\
    Hubbard model with boundary fields}

  \author[Chemnitz,PortoAlegre]{X.-W.~Guan},
  \author[PortoAlegre]{A.~Foerster}, \author[MiltonKeynes]{U.~Grimm},
  \author[Chemnitz]{R.A.~R\"{o}mer}, \author[Chemnitz]{M.~Schreiber}
  \address[Chemnitz]{Institut f\"ur Physik, Technische Universit\"at,
    09107 Chemnitz, Germany} \address[PortoAlegre]{Instituto de Fisica
    da UFRGS, Av.\ Bento Goncalves, 9500,\\ Porto Alegre, 91501-970,
    Brazil} \address[MiltonKeynes]{Applied Mathematics Department,
    Faculty of Mathematics and Computing,\\ The Open University,
    Walton Hall, Milton Keynes MK7 6AA, U.K.}

\begin{abstract}
  A strongly correlated electron system associated with the quantum
  superalgebra $\mbox{U}_q[\mbox{osp}(2|2)]$ is studied in the
  framework of the quantum inverse scattering method. By solving the
  graded reflection equation, two classes of boundary-reflection
  $K$-matrices leading to four kinds of possible boundary interaction
  terms are found. Performing the algebraic Bethe ansatz, we
  diagonalize the two-level transfer matrices which characterize the
  charge and the spin degrees of freedom, respectively. The
  Bethe-ansatz equations, the eigenvalues of the transfer matrices and
  the energy spectrum are presented explicitly. We also construct two
  impurities coupled to the boundaries. In the thermodynamic limit,
  the ground state properties and impurity effects are discussed.
\end{abstract}

\end{frontmatter}

\section{Introduction}
\label{sec1}
In recent decades, there has been considerable interest in the study
of strongly correlated electron systems in reduced dimensions
exhibiting non-Fermi-liquid behaviour \cite{SutR93,nfl1,nfl2}. This has been
motivated by the surprising properties of high-$T_{c}$ superconductors
\cite{HTc}, heavy-fermion alloys and compounds \cite{HF}, and the concept of
Luttinger liquids \cite{Lutc,Schulz,Mattis-Lieb}. Similar non-standard behaviour that lies
outside the realm of Fermi liquid theory was also observed in the
magnetic properties of systems displaying the Kondo effect
\cite{Kondo}. The Kondo problem was solved exactly by means of the
Bethe ansatz and the quantum inverse scattering method (QISM)
\cite{QISM}. Following this approach, many one-dimensional integrable
quantum systems with impurities \cite{imp1} have been constructed as
inhomogeneous solutions of the Yang-Baxter equations (YBE) \cite{YBE}.

A particularly intriguing situation corresponds to perfect
backscattering impurities which can be realized conveniently by
integrable open boundary conditions (BC). A systematic approach to
handle quantum systems with backscattering boundaries is provided by
Sklyanin's work on the reflection equations (RE) \cite{EK}. In analogy
with the YBE in the bulk, the RE guarantee the factorization of the
scattering matrices at the boundaries. Thus, starting from a solution of
the YBE, which yields a solvable model with periodic BC, one constructs
suitable integrable boundary conditions such that the RE are fulfilled
\cite{open1,open2,open3,open4}. Due to integrability, the bulk
impurities obtained by inhomogeneous solutions of the YBE are pure
forward scatterers. Thus their combination with the backscattering
boundaries may be expected to model physically relevant impurity
systems.  In the context of boundary integrable quantum field theories
\cite{bqft}, a model with impurities can be mapped onto a model with
certain boundary conditions. Because impurity effects play a decisive
role for the transport in quantum wires, general boundary conditions for
strongly correlated electron systems open many opportunities to
investigate the transport properties in Luttinger liquids or quantum
wires. A proper boundary field may have a feasible realization by
applying boundary external voltages and external magnetic fields in
experiments on quantum wires \cite{q-w}. Moreover, we may expect that
the local state induced by the boundary fields inherits signatures of
the bulk Luttinger liquid \cite{Lutc,Schulz,Mattis-Lieb}. The physical
quantities such as magnetization, the compressibility, susceptibility,
and specific heat, etc., may be manipulated by the Bethe-ansatz
equations. For this reason, integrable models that combine bulk and
boundary impurities have recently attracted much attention
\cite{imp2,imp3,imp4,Zv}.

Of particular interest are strongly correlated electron systems
associated with {\em supersymmetric} solutions of the YBE.  Models
corresponding to non-excep\-tio\-nal Lie superalgebras as, for
instance, $\mbox{gl}(2|1)$ and $\mbox{osp}(2|2)$
\cite{supal,deg,doub}, have provided interesting non-perturbative
information \cite{sces,Essler} for generalizations of well-known
models such as, e.g., the Hubbard model \cite{LW}. A further
generalization was achieved by considering the solution of the YBE
related to the quantum superalgebra $\mbox{U}_q[\mbox{osp}(2|2)]$
\cite{deg}. This model has two fermionic and two bosonic degrees of
freedom. Furthermore, the Hamiltonian \cite{doub} corresponds to a
lattice regularization of the integrable double sine-Gordon model
\cite{doubS}. The continuum version of this model with boundary fields
is known to describe tunneling effects in quantum wires \cite{wire}.
These wires are believed to represent a realization of Luttinger
liquids. We remark that the coordinate Bethe ansatz for an open
$\mbox{U}_q[\mbox{osp}(2|2)]$ chain has recently been studied
\cite{MG1}.

In the present work, we perform the algebraic Bethe ansatz for a
supersymmetric open chain associated with the quantum superalgebra
$\mbox{U}_q[\mbox{osp}(2|2)]$. Due to the supersymmetric structure of
the model, we use the graded version of the QISM
\cite{Essler}. Starting from the $\mbox{U}_q[\mbox{osp}(2|2)]$
solution of the graded YBE, we construct a supersymmetric correlated
electron system with boundary fields by solving the graded RE. We find
that the model contains a hidden anisotropic XXZ open chain
characterizing the spin degrees of freedom \cite{SutR93}. This plays a
crucial role in our solution, which proceeds in two steps, treating
charge and spin degrees of freedom separately by a nested graded Bethe
ansatz. This structure also suggests a natural way to incorporate
impurities coupling to charge and spin degrees of freedom while
preserving the integrability of the model. From the Bethe-ansatz
equations, we obtain the ground state energy at half-filling in the
thermodynamic limit.  We also discuss integrable impurities coupled to
the boundaries.

The paper is organized as follows. In section \ref{sec2}, we present
the $\mbox{U}_q[\mbox{osp}(2|2)]$ solution of the YBE and solve the
corresponding RE. Furthermore, we give an explicit expression of the
Hamiltonian in terms of fermionic operators. Section \ref{sec3} is
devoted to the derivation and the solution of the Bethe-ansatz
equations by means of the graded QISM. In section \ref{sec4}, the
ground state energy in the thermodynamic limit is obtained for the model
with boundary fields and impurities. We conclude in section \ref{sec5}.

\section{The solutions of the graded RE}
\label{sec2}
We begin by considering a two dimensional classical lattice model,
where, to each bond of the lattice, two bosonic and two fermionic
degrees of freedom are associated. It is a vertex model, hence
interaction takes place at each vertex, and the energies of the
various local configurations determine the statistical weight of the
configuration. The corresponding Boltzmann weights form a $2^4\times
2^4$ quantum $R$-matrix with grading `bffb', where `b' stands for {\em
bosonic} and `f' for {\em fermionic}. It has the form
\begin{equation}
\hspace*{-\mathindent}
R_{12}(\lambda)  =
\mbox{\begin{footnotesize}$\left(\matrix{
w_1&0&0&0&0&0&0&0&0&0&0&0&0&0&0&0\cr
0&\!\!w_3&0&0&\!\!w_4&0&0&0&0&0&0&0&0&0&0&0\cr
0&0&\!\!w_3&0&0&0&0&0&\!\!w_4&0&0&0&0&0&0&0\cr
0&0&0&\!\!w_7&0&0&\!\!\!\!\!w_{10}\!\!\!&0&0&
\!\!\!\!\!w_{10}\!\!\!&0&0&\!\!w_{11}\!\!\!&0&0&0\cr
0&\!\!w_2&0&0&\!\!w_3&0&0&0&0&0&0&0&0&0&0&0\cr
0&0&0&0&0&\!\!\!\!-1&0&0&0&0&0&0&0&0&0&0\cr
0&0&0&\!\!w_6&0&0&\!\!\!\!\!w_9\!&0&0&
\!\!\!\!\!w_8\!&0&0&\!\!\!\!\!w_{10}\!\!\!&0&0&0\cr
0&0&0&0&0&0&0&\!\!w_3&0&0&0&0&0&\!\!w_4&0&0\cr
0&0&\!\!w_2&0&0&0&0&0&\!\!w_3&0&0&0&0&0&0&0\cr
0&0&0&\!\!w_6&0&0&\!\!\!\!\!w_8\!&0&0&\!\!\!\!\!w_9\!&0&0&
\!\!\!\!\!w_{10}\!\!\!&0&0&0\cr
0&0&0&0&0&0&0&0&0&0&\!\!\!\!-1&0&0&0&0&0\cr
0&0&0&0&0&0&0&0&0&0&0&\!\!w_3&0&0&\!\!w_4&0\cr
0&0&0&\!\!w_5&0&0&\!\!w_6&0&0&\!\!w_6&0&0&\!\!w_7&0&0&0\cr
0&0&0&0&0&0&0&\!\!w_2&0&0&0&0&0&\!\!w_3&0&0\cr
0&0&0&0&0&0&0&0&0&0&0&\!\!w_2&0&0&\!\!w_3&0\cr
0&0&0&0&0&0&0&0&0&0&0&0&0&0&0&\!\!w_1}\right)$\end{footnotesize}}
\nonumber\\* \label{R12}
\end{equation}
where
\begin{eqnarray}
w_1 & = & \frac{x-q^2}{1-x\,q^2},\qquad\qquad\quad
w_2  =\frac{1-q^2}{1-x\,q^2},\qquad
w_3  = \frac{q(x-1)}{1-x\,q^2},\nonumber\\
w_4 & = &x\,w_2,\qquad\qquad\qquad\quad
w_5  = \frac{2\,w_2}{1+x},\qquad\quad
w_6  = \frac{1-x}{1+x}\,w_2,\nonumber\\
w_7 & = &-\frac{1+q^2\,x}{1-q^2}\,w_6,\qquad\quad
w_8 = -\frac{2x\,w_2}{1+x},\qquad
w_9  = -\frac{x+q^2}{1-q^2}\,w_6,\nonumber\\
w_{10} & =& x\,w_6,\qquad\qquad\qquad\quad
w_{11}  = \frac{2x^2\,w_2}{1+x},\label{eq:w}
\end{eqnarray}
with $x=\e^{\lambda}$ characterizing the difference $\lambda$ of the
pseudo-momenta of the particles whose two-body scattering is described
by the quantum $R$-matrix (\ref{R12}); $q$ is the deformation parameter.

In what follows, we are going to use also the standard diagrammatic
representation \cite{QISM}, where the $R$-matrix corresponds to the two-particle
scattering picture
\begin{equation}
R_{12}(\lambda)  =
\begin{array}{c}
\unitlength=0.50mm
\begin{picture}(20.,25.)
\put(20.,0){\vector(-1,1){20.}}
\put(0.,0){\vector(1,1){20.}}
\put(0.,0.){\line(1,1){20.}}
\put(0.,20.){\line(1,-1){20.}}
\end{picture}
\end{array}.
\end{equation}
The quantum $R$-matrix (\ref{R12}) satisfies the graded YBE
\begin{equation}
R_{12}(\lambda-\nu)R_{13}(\lambda)R_{23}(\nu)
= R_{23}(\nu)R_{13}(\lambda )R_{12}(\lambda -\nu)
\label{YBE}
\end{equation}
guaranteeing the integrability of the model with periodic BC, i.e.,
the factorization of the scattering matrices into two-body scattering
matrices \cite{YBE}. Above, $R_{ij}$ with $i=1,2$ and $j=2,3$, denotes
on which of the $i$th and $j$th spaces of $V_1\otimes _sV_2\otimes
_sV_3$ the $R$-matrix acts. In the remaining space, $R_{ij}$ acts an
identity.  Here ${\otimes}_{S}$ denotes the graded tensor product
\begin{equation}
[A{\otimes }_{S}B]_{\alpha\beta, \gamma
\delta}=(-1)^{[P(\alpha)+P(\gamma)]P(\beta)}
A_{\alpha\gamma}B_{\beta\delta},
\end{equation}
with the Grassmann parities
obey $P(1)=P(4)=0$ and $P(2)=P(3)=1$ with respect to the grading
`bffb'.

For other BC such as twisted and open BC, the graded YBE will still
account for the bulk part of the model, but the boundary terms have to
be chosen appropriately in order to preserve the integrability of the
model. In particular, solutions of the RE yield integrable vertex
models with open (reflecting) boundaries or the equivalent integrable
quantum spin chains with boundary fields.  In the context of two-body
scattering, the RE characterize the consistency conditions for the
factorizations of the two-body boundary scattering matrices at
boundaries.  Taking into account the grading `bffb', the left and right reflection matrices,
$K_{-}$ and $K_{+}$,  are required to satisfy the following RE
\begin{eqnarray}
\lefteqn{
 R_{12}(\lambda-\nu)
 \stackrel{1}{K}_{-}\!(\lambda)\,
 R_{21}(\lambda+\nu )
 \stackrel{2}{K}_{-}\!(\nu )} \nonumber\\
& = &
 \stackrel{2}{K}_{-}\!(\nu )\,
 R_{12}(\lambda +\nu )
 \stackrel{1}{K}_{-}\!(\lambda)\,
 R_{21}(\lambda-\nu ),  \label{RE1}\\
\lefteqn{
 R_{21}^{{\rm St}_{1}\overline{{\rm St}}_2}(\nu -\lambda)
 \stackrel{1}{K}\!\mbox{}_{+}^{{\rm St}_1}(\lambda)\,
 R_{12}^{{\rm St}_1\overline{{\rm St}}_2}(-\lambda-\nu)
 \stackrel{2}{K}\!\mbox{}_{+}^{\overline{{\rm St}}_2}(\nu )}
 \nonumber \\
& = &
 \stackrel{2}{K}\!\mbox{}_{+}^{\overline{{\rm St}}_2}(\nu )\,
 R_{21}^{{\rm St}_1 \overline{{\rm St}}_2}(-\lambda-\nu)
 \stackrel{1}{K}\!\mbox{}_{+}^{{\rm St}_1}\!(\lambda)\,
 R_{12}^{{\rm St}_1\overline{{\rm St}}_2}(\nu -\lambda),
\label{RE2}
\end{eqnarray}
respectively.
Here, we used the conventional notation
\begin{equation}
\stackrel{1}{X} \equiv X\otimes_{S}\mathbf{I}_{V_2},\qquad
\stackrel{2}{X} \equiv \mathbf{I}_{V_1}\otimes_{S}X,
\end{equation}
where $\mathbf{I}_{V}$ denotes the identity operator on $V$,
and, as usual, $R_{21}={\bf P}\cdot R_{12}\cdot {\bf P}$.
Here ${\bf P}$ is the graded permutation operator which can be represented by a
$2^4\times 2^4$ matrix, i.e.,
\begin{equation}
{\bf P}_{\alpha\beta, \gamma
\delta}=(-1)^{P(\alpha)P(\beta)}
\delta_{\alpha\delta}\delta_{\beta\gamma}.
\end{equation}
Furthermore, superscripts ${\rm St}_a$ and
$\overline{{\rm St}}_{a}$ denote the supertransposition in the space
with index $a$ and its inverse, respectively,
\begin{equation}
(A_{ij})^{{\rm St}}=(-1)^{[P(i)+P(j)]P(i)}A_{ji},
\qquad
(A_{ij})^{\overline{{\rm St}}} =(-1)^{[P(i)+P(j)]P(j)}A_{ji}.
\end{equation}
We found two solutions of the RE (\ref{RE1}) and (\ref{RE2}) for
diagonal boundary $K_{\pm}$-matrices
\begin{eqnarray}
K_{-}(\lambda) & = &
\unitlength=0.50mm
\begin{picture}(20.,25.)
\put(17.3,0.){\oval(15.,15.)[l]}
\multiput(10.,-10.)(0,2){10}{\line(0,1){1.0}}
\put(10.00,0.){\circle{1.5}}
\put(17.1,7.4){\vector(1,0){0.3}}
\end{picture}~~
 =
\left(\matrix{K^{(1)}_{-}(\lambda)&0&0&0\cr
              0&K^{(2)}_{-}(\lambda)&0&0\cr
              0&0&K^{(3)}_{-}(\lambda)&0\cr
              0&0&0&K^{(4)}_{-}(\lambda)\cr}\right), \label{K-M} \\
K_+(\lambda) & = & ~~
\unitlength=0.50mm
\begin{picture}(20.,25.)
\multiput(10.,-10.)(0,2){10}{\line(0,1){1.0}}
\put(2.2,0.){\oval(15.,15.)[r]}
\put(10.00,0.){\circle{1.5}}
\put(2.1,-7.5){\vector(-1,0){0.3}}
\end{picture}
=
K^{{\rm St}}_-(\lambda\rightarrow -\lambda-\mathrm{i}\pi ,\,
\xi _-\rightarrow -\frac{1}{\xi _+}).
\label{Kp}
\end{eqnarray}
In the first solution, the entries are given as
\begin{eqnarray}
K^{(1)}_-(\lambda) &=& (x+\xi _-q)(x-\xi _-q^{-1}),
\nonumber\\
K^{(2)}_-(\lambda) &=& K^{(3)}_-(\lambda)= (x^{-1}+\xi _-q)(x -\xi _-q^{-1}),
\nonumber\\
K^{(4)}_-(\lambda) &=& (x^{-1}+\xi _-q)(x^{-1}-\xi _-q^{-1}),
\label{Km}
\end{eqnarray}
whereas the second solutions corresponds to
\begin{eqnarray}
K^{(1)}_{-}(\lambda) &=&
 (x+\xi_{-}q)
 (x-\xi_{-}^{-1}q)
 (x-\xi_{-}q^{-1}), \nonumber\\
K^{(2)}_{-}(\lambda) &=&
 (x^{-1}+\xi_{-}q)
 (x-\xi_{-}^{-1}q)
 (x -\xi_{-}q^{-1}), \nonumber \\
K^{(3)}_{-}(\lambda) &=&
 (x+\xi_{-}q)
 (x^{-1}-\xi_{-}^{-1}q)
 (x -\xi_{-}q^{-1}), \nonumber\\
K^{(4)}_{-}(\lambda) &=&
 (x+\xi_{-}q)
 (x^{-1}-\xi_{-}^{-1}q)
 (x^{-1}-\xi_{-}q^{-1}). \label{Km2}
\end{eqnarray}
Here $x=\e^{\lambda }$ as defined in (\ref{eq:w}) and $\xi _{\pm}$ are
free parameters characterizing the boundary fields and the boundary
interactions.  We emphasize that the permutation-transposition
symmetry
\begin{equation}
R_{12}^{{\rm St}_{1}\overline{{\rm St}}_{2}}(\lambda )={\bf P} R_{12}(\lambda){\bf P},
\end{equation}
the unitarity
\begin{equation}
R_{12}(\lambda)R_{21}(-\lambda)=\mathbf{I},
\label{uni}
\end{equation}
and the graded crossing symmetry
\begin{equation}
R_{12}(\lambda)=\,
\stackrel{1}{V}\!
R_{12}^{{\rm St}_2}(-\lambda-\mathrm{i}\pi)
\stackrel{1}{V}\!\mbox{}^{-1}\,
\frac{\zeta(-\lambda-\mathrm{i}\pi)}{\zeta(\lambda)},
\label{eq-gcs}
\end{equation}
with
\begin{equation}
V=\left(\matrix{0&0&0&\!\!\!-1\cr
                0&0&1&0\cr
                0&1&0&0\cr
                1&0&0&0}\right),\qquad
\zeta (\lambda)=\frac{(1-xq^2)}{(1-x)},
\end{equation}
not only result in the isomorphism (\ref{Kp}) between $K_{+}(\lambda)$
and $K_{-}(\lambda)$, but also constitute the necessary ingredients
for the integrability of the model with boundaries.  We also remark
that the grading does not play a role for the diagonal
$K_-(\lambda)$-matrix (\ref{Km}), which coincides with the one for the
non-graded $R$-matrix given in Ref.~\cite{MG1}. However, its companion
$K_+(\lambda)$ given by (\ref{Kp}) is different from that in the
non-graded case because it has to obey the graded crossing symmetry
(\ref{eq-gcs}).  It is worth emphasizing that the boundary parameters
$\xi_{-}$ and $\xi_{+}$ should inherit the same crossing property as
that imposed on the pseudo-momenta $\lambda$ such that the boundary
terms of the corresponding Hamiltonian are completely symmetric. In
Appendix \ref{sec-boundary-matrices} we present an ansatz to work out
the solutions (\ref{Km}) as well as (\ref{Km2}).

A more important object in the context of the QISM is the transfer
matrix of an integrable system, which can be considered as a
generating function of the infinite integrals of motion due to its
commutativity for different values of the spectral parameter.
Actually, the RE (\ref{RE1}) and (\ref{RE2}) together with the YBE
(\ref{YBE}) and the symmetries of the quantum $R$-matrix ensure the
commutativity of the double-row transfer matrix
\begin{eqnarray}
\tau (\lambda)& = &~~~~~
\begin{array}{c}
\unitlength=0.50mm
\begin{picture}(95.,49.)
\put(15.,25.){\oval(20.,20.)[l]}
\put(75.,55.){\makebox(0,0)[cc]{$L$}}
\put(30.,55.){\makebox(0,0)[cc]{$2$}}
\put(20.,55.){\makebox(0,0)[cc]{$1$}}
\put(65.,0.){\vector(0,1){45.}}
\put(75.,0.){\vector(0,1){45.}}
\put(30.,0.){\vector(0,1){45.}}
\put(20.,0.){\vector(0,1){45.}}
\put(73.,35.){\vector(1,0){8.}}
\put(57.,15.){\line(1,0){25.}}
\put(47.,35.){\makebox(0,0)[cc]{$\cdots$}}
\put(47.,15.){\makebox(0,0)[cc]{$\cdots$}}
\put(15.,15.){\line(1,0){21.}}
\multiput(5.,0.)(0,4){12}{\line(0,1){2.0}}
\put(5.00,25.){\circle{1.5}}
\put(80.,25.){\oval(20.,20.)[r]}
\put(90.,25.){\circle{1.5}}
\multiput(90.,0.)(0,4){12}{\line(0,1){2.0}}
\put(30.,15.){\vector(-1,0){16.}}
\put(57.,35.){\line(1,0){25.}}
\put(15.,35.){\line(1,0){21.}}
\end{picture}
\end{array}\,\nonumber \\
& = &
\; {\rm Str}_0\left[{K}_+(\lambda)T(\lambda)K_-(\lambda)
T^{-1}(-\lambda)\right].\label{TM}
\end{eqnarray}
Here ${\rm Str}_{0}$ denotes the supertrace carried out in auxiliary
space $V_{0}$. The monodromy matrices $T(\lambda)$ and
$T^{-1}(-\lambda)$ are defined by
\begin{eqnarray}
T(\lambda )& =&
\begin{array}{c}
\unitlength=0.50mm
\begin{picture}(99.,29.)
\put(64.,15.){\line(1,0){24.}}
\put(55.,15.){\makebox(0,0)[cc]{$\cdots$}}
\put(47.,15.){\vector(-1,0){31.}}
\put(30.,0.){\vector(0,1){28.}}
\put(40.,0.){\vector(0,1){28.}}
\put(70.,0.){\vector(0,1){28.}}
\put(80.,0.){\vector(0,1){28.}}
\put(80.,35.){\makebox(0,0)[cc]{$L$}}
\put(40.,35.){\makebox(0,0)[cc]{$2$}}
\put(30.,35.){\makebox(0,0)[cc]{$1$}}
\end{picture}
\end{array}\nonumber
\\ &=&
R_{L,0}(\lambda )R_{L-1,0}(\lambda )\cdots R_{2,0}(\lambda )R_{1,0}(\lambda ),\label{tm}
\end{eqnarray}
\vspace{1.5cm}
\begin{eqnarray}
T^{-1}(-\lambda )& =&
\begin{array}{c}
\unitlength=0.50mm
\begin{picture}(99.,29.)
\put(64.,15.){\vector(1,0){24.}}
\put(55.,15.){\makebox(0,0)[cc]{$\cdots$}}
\put(47.,15.){\line(-1,0){31.}}
\put(30.,0.){\vector(0,1){28.}}
\put(40.,0.){\vector(0,1){28.}}
\put(70.,0.){\vector(0,1){28.}}
\put(80.,0.){\vector(0,1){28.}}
\put(80.,35.){\makebox(0,0)[cc]{$L$}}
\put(40.,35.){\makebox(0,0)[cc]{$2$}}
\put(30.,35.){\makebox(0,0)[cc]{$1$}}
\end{picture}
\end{array}  \nonumber \\
&= &
R_{0,1}(\lambda )R_{0,2}(\lambda )\cdots R_{0,L-1}(\lambda )R_{0,L}(\lambda ),\label{tmi}
\end{eqnarray}
respectively.  The Hamiltonian associated with the quantum $R$-matrix
(\ref{R12}) is related to the double-row transfer matrix (\ref{TM})
as
\begin{equation}
\tau (\lambda)=c_{1}\lambda+c_{2}(H+{\rm const.})\lambda^{2}+
\ldots, \label{HTR}
\end{equation}
where $c_{i}$, $i=1,2$, are scalar functions of the boundary
parameters $\xi_{\pm}$. Taking into account the Grassmann parity of
the host vertices, after some lengthy algebra, one can present the
Hamiltonian explicitly in terms of the fermionic creation and
annihilation operators $c^{\dagger}_{j,\sigma}$ and $c_{j,\sigma}^{}$
acting on the site $j$ and carrying the spin index $\sigma =\pm $. It
is given by
\begin{eqnarray}
H  & = &
 \sum _{j=1}^{L-1}H_{j,j+1}+U(\xi _-)n_{1,+}n_{1,-}+U(\xi _+)n_{L,+}n_{L,-}
\nonumber\\
 & & + \sum _{\sigma =\pm}(V_{1,\sigma }\,n_{1,\sigma}+
 V_{L,\sigma }\,n_{L,\sigma}),\label{Ham}
\end{eqnarray}
where $n_{j,\sigma}=c^{\dagger}_{j,\sigma}c_{j,\sigma}^{}$ is
the fermion number operator, and  the bulk Hamiltonian  is chosen
as
\begin{eqnarray}
H_{j,j+1} & = &
2\sum_{\sigma=\pm} \left[ c_{j,\sigma}^{\dagger}
c_{j+1, \sigma} + \rm{h.c.}  \right ]
\left[ 1-n_{j,-\sigma }-n_{j+1,-\sigma }+n_{j,-\sigma }n_{j+1,-\sigma }\right]\nonumber\\
& &
-(q-q^{-1})\sum_{\sigma=\pm} \left[ c_{j, \sigma}^{\dagger}
c_{j+1, \sigma} - {\rm h.c.}  \right ]
\left[n_{j+1,-\sigma }-n_{j,-\sigma }\right] \nonumber \\
& &
+(q+q^{-1})\left[ c_{j,+}^{\dagger} c_{j,-}^{\dagger}
c_{j+1,-} c_{j+1,+}-c_{j,+}^{\dagger} c_{j+1,-}^{\dagger} c_{j+1,+} c_{j,-} +
{\rm h.c.} \right]\nonumber \\
& &
+(q+q^{-1})\left [ n_{j,+} n_{j,-} +n_{j+1,+} n_{j+1,-}
 + n_{j,+} n_{j+1,-}\right.\nonumber\\
& &\hphantom{(q+q^{-1})\left[\;\right. }\left.
+n_{j,-}n_{j+1,+} -n_{j} -n_{j+1} \right ]\nonumber\\
& &
+2(q+q^{-1})\mathbf{I},\label{Hambulk}
\end{eqnarray}
with $n_j=n_{j,+}+n_{j,-}$ and $\mathbf{I}$ the identity operator. The
remaining terms in (\ref{Ham}) are the on-site Coulomb coupling
$U(\xi_{\pm})$ and the chemical potentials $V_{1,\sigma}$ and
$V_{L,\sigma}$ at the ends of the chain. For the first
solution (\ref{Km}), we find
\begin{eqnarray}
U(\xi_{\pm}) & = &
\frac{2(q^2-q^{-2})q\xi_{\pm }}{(q-\xi _{\pm})(1+q\xi _{\pm}) },
\label{eq:u}\\
V_{1,\sigma}& = &-\frac{(q-q^{-1})(q+\xi _-)}{(q-\xi _-)},\quad
V_{L,\sigma }=-\frac{(q-q^{-1})(q+\xi _+)}{(q-\xi _+)};\label{bt1}
\end{eqnarray}
whereas the second solution (\ref{Km2}) yields the same expression
(\ref{eq:u}) for $U(\xi_{\pm})$, $V_{1,-}$ and $V_{L,-}$ but
\begin{equation}
V_{1,+} = -\frac{(q-q^{-1})(q\xi _--1)}{(1+q\xi _-)},\quad
V_{L,+} = -\frac{(q-q^{-1})(q\xi _+-1)}{(1+q\xi _+)},\label{bt2}
\end{equation}
as presented in Appendices \ref{sec-boundary-matrices} and
\ref{sec-hamiltonian}. The Hamiltonian (\ref{Ham}) contains hopping
terms with occupation numbers, double hopping terms, on- and off-site
Coulomb interaction in the bulk, as well as on-site Coulomb
interaction and chemical potentials at the boundaries. We notice that
the boundary terms corresponding to the first solution (\ref{Km}) act
as boundary chemical potentials, whereas in the second case (\ref{Km2}) they act
as boundary magnetic fields.  The two cases could provide four
possible classes of boundary conditions, which lead to different
boundary shift factors in the Bethe-ansatz equations of the model. In
order to keep the Hamiltonian hermitian, we restrict ourselves to
$q=\e^{\mathrm{i}\gamma}$ and boundary parameters
$\xi_{\pm}=\e^{\mathrm{i}\xi^{\pm}}$ with real $\gamma$ and
$\xi^{\pm}$.  We note that the Hamiltonian in Ref.~\cite{doub} differs
from the bulk part (\ref{Hambulk}), but that they are related to each
other by a canonical transformation.  Nevertheless, it will be shown
that this transformation does not change the Bethe-ansatz equations in the bulk.

So far, we finished the first step towards the algebraic Bethe-ansatz
solution for the model associated with the quantum $R$-matrix
(\ref{R12}). Next we shall proceed with the factorization of the
transfer matrix (\ref{TM}). In this paper, we restrict ourselves to
the first solution (\ref{Km}) of the RE, which leads to
a perfect factorization of the transfer matrix (\ref{TM}) acting on the
pseudo-vacuum state. The second class of solutions of the RE, given in
(\ref{Km2}), leads to a very complicated factorization of
the transfer matrix (\ref{TM}) acting on  the pseudo-vacuum
state and the multi-particle states. However the ansatz formulated in the
following section  works in a similar way for the model with other
boundary terms.

\section{The algebraic Bethe-ansatz approach}
\label{sec3}

In order to accomplish the algebraic Bethe ansatz for an integrable
system with boundaries, we first need to diagonalize the transfer
matrix of the model acting on the pseudo-vacuum state. As usual, we
rewrite the transfer matrix (\ref{TM}) in the following form
\begin{equation}
\tau (\lambda)={\rm Str}_0\left[ K_+(\lambda)U_-(\lambda)\right],
\label{TM2}
\end{equation}
where $U_-(\lambda)$ is defined by
\begin{equation}
U_-(\lambda ) = ~~~
\begin{array}{c}
\unitlength=0.50mm
\begin{picture}(85.,55.)
\multiput(5.,0.)(0,4){12}{\line(0,1){2.0}}
\put(45.,15.){\makebox(0,0)[cc]{$\cdots$}}
\put(15.,25.){\oval(20.,20.)[l]}
\put(75.,55.){\makebox(0,0)[cc]{$L$}}
\put(30.,55.){\makebox(0,0)[cc]{$2$}}
\put(20.,55.){\makebox(0,0)[cc]{$1$}}
\put(75.,0.){\vector(0,1){45.}}
\put(65.,0.){\vector(0,1){45.}}
\put(30.,0.){\vector(0,1){45.}}
\put(20.,0.){\vector(0,1){45.}}
\put(73.,35.){\vector(1,0){10.}}
\put(57.,35.){\line(1,0){25.}}
\put(57.,15.){\line(1,0){25.}}
\put(45.,35.){\makebox(0,0)[cc]{$\cdots$}}
\put(15.,15.){\line(1,0){21.}}
\put(15.,35.){\line(1,0){21.}}
\put(5.,25.){\circle{1.5}}
\end{picture}
\end{array}
=
T(\lambda )K_-(\lambda )T^{-1}(-\lambda ).
\end{equation}
One can verify that  $U_-(\lambda)$ also
satisfies the graded RE (\ref{RE1}). With regard to the structure of the
$R$-matrix (\ref{R12}), we choose the standard ferromagnetic
pseudo-vacuum state \cite{Mar}
\begin{equation}
|0\rangle =|0\rangle _L\otimes \cdots\otimes|0\rangle _i\otimes\cdots
\otimes |0\rangle _1,\label{pseudo-vacuum}
\end{equation}
where $ |0\rangle _i= \left(\matrix{1\cr 0}\right)_i \otimes
\left(\matrix{1\cr 0}\right)_i $ acts as a highest-weight
vector. Following Refs.\ \cite{Mar,G1}, we label the elements of the
monodromy matrix $T(\lambda)$ by
\begin{eqnarray}
T(\lambda) & = &
\left(\matrix{B(\lambda)&B_1(\lambda)&B_2(\lambda)&F(\lambda)\cr
C_1(\lambda)&A_{11}(\lambda)&A_{12}(\lambda)&E_1(\lambda)\cr
C_2(\lambda)&A_{21}(\lambda)&A_{22}(\lambda)&E_2(\lambda)\cr
C_3(\lambda)&C_4(\lambda)&C_5(\lambda)&D(\lambda)\cr }\right),
\end{eqnarray}
and further
\begin{eqnarray}
T^{-1}(-\lambda) & = &
\left(\matrix{\bar{B}(\lambda)&
\bar{B}_1(\lambda)&\bar{B}_2(\lambda)&\bar{F}(\lambda)\cr
\bar{C}_1(\lambda)&\bar{A}_{11}(\lambda)&\bar{A}_{12}(\lambda)&
\bar{E}_1(\lambda)\cr
\bar{C}_2(\lambda)&\bar{A}_{21}(\lambda)&\bar{A}_{22}(\lambda)&
\bar{E}_2(\lambda)\cr \bar{
C}_3(\lambda)&\bar{C}_4(\lambda)&\bar{C}_5(\lambda)&
\bar{D}(\lambda)\cr }
\right), \label{TIM} \\[1ex]
U_-(\lambda) & = &  \left(\matrix{\tilde{B}(\lambda)&
\tilde{B}_1(\lambda)&\tilde{B}_2(\lambda)&\tilde{F}(\lambda)\cr
\tilde{C}_1(\lambda)&\tilde{A}_{11}(\lambda)&\tilde{A}_{12}(\lambda)&
\tilde{E}_1(\lambda)\cr
\tilde{C}_2(\lambda)&\tilde{A}_{21}(\lambda)&\tilde{A}_{22}(\lambda)&
\tilde{E}_2(\lambda)\cr
\tilde{C}_3(\lambda)&\tilde{C}_4(\lambda)&\tilde{C}_5(\lambda)&
\tilde{D}(\lambda)\cr }\right).\label{UM}
\end{eqnarray}
{}From the structure of the $R$-matrix (\ref{R12}),  the relation
(\ref{uni}) --- which one uses to construct the inverse $R$-matrix ---
and the definitions (\ref{tm}) and (\ref{tmi}), one can deduce that
the operators $B_a(\lambda)$ and $\bar{B}_a(\lambda)$ ($a=1,2$) act as
creation fields acting on the reference state, creating particles with
pseudo-momenta $\lambda$ and $-\lambda$, respectively. While
$E_a(\lambda)$ and $\bar{E}_a(\lambda)$ are the `dual' creation fields
to  $B_a(\lambda)$ and $\bar{B}_a(\lambda)$, the
operators $C_i(\lambda)$ and $\bar{C}_i(\lambda)$ ($i=1,\cdots,5$) behave as
annihilation fields.  Furthermore, using an invariant of the
Yang-Baxter algebra
\begin{equation}
\stackrel{2}{T}\mbox{$\!$}^{-1}(-\lambda)
 R_{12}(2\lambda )
 \stackrel{1}{T}\!(\lambda) =
 \stackrel{1}{T}\!(\lambda)
 R_{12}(2\lambda )
 \stackrel{2}{T}\mbox{$\!$}^{-1}(-\lambda ),\label{YBA}
\end{equation}
we obtain, apart from an overall factor
$Q(\lambda)=K^{(1)}_{-}(\lambda)K^{(1)}_{+}(\lambda)$, the eigenvalue
of the transfer matrix as
\begin{equation}
\tau(\lambda)|0\rangle  = \left\{ W^{+}_{1}(\lambda)\tilde{B}(\lambda)
+\sum_{a=1}^{2}W^{+}_{a+1}(\lambda)\hat{A}_{aa}(\lambda)
+W^{+}_{4}(\lambda)\hat{D}(\lambda)\right\}|0\rangle,
\label{fact}
\end{equation}
where we introduced the transformations
\begin{eqnarray}
\hat{A}_{aa}(\lambda) &=&
\tilde{A}_{aa}(\lambda)-\frac{q^2-1}{q^2-x^2}\tilde{B}(\lambda)
=W_{a+1}^-(\lambda)\bar{A}_{aa}(\lambda)A_{aa}(\lambda), \label{transform1}\\
\hat{D}(\lambda) &=&\tilde{D}(\lambda)-\frac{q^2-1}{x^2q^2-1}\sum _{a=1}^{2}\hat{A}_{aa}(\lambda)
-\frac{2(q^2-1)}{(q^2-x^2)(x^2+1)}\tilde{B}(\lambda)\nonumber\\
&=&W_4^-(\lambda)\bar{D}(\lambda)D(\lambda). \label{transform2}
\end{eqnarray}
Here,
\begin{eqnarray}
W^-_1(\lambda)& = & 1,\\
W^+_1(\lambda)& = &
\frac{(x^2+q^2)(x^2-1)
      (x+q\xi_+)(xq-\xi _+)}
     {(x^2-q^2)(x^2+1)
      (x-q\xi_+)(xq+\xi _+)},\\
W^+_{2}(\lambda)& = &W^+_{3}(\lambda)=
-\frac{xq(x^2-1)(x\xi_++q)
       (x q-\xi _+)}
      {(x^2q^2-1)(x-q\xi_+)(xq+\xi _+)},\\
W^-_{2}(\lambda)& = &W^-_{3}(\lambda)=
\frac{q(x^2-1)(x\xi _-+q)}
     {x(x^2-q^2)(x+q\xi _-)},\\
W^+_4(\lambda)& = &
-\frac{x^2(xq\xi_+-1)(x\xi _++q)}
      {(x-q\xi _+)(xq+\xi_+)},\\
W^-_4(\lambda)& = &
-\frac{(x^2-1)(x^2q^2+1)
       (xq\xi_--1)(x\xi_-+q)}
      {x^2(x^2+1)(x^2q^2-1)(xq-\xi_-)
       (x+q\xi_-)},\\
\tilde{B}(\lambda)|0\rangle & = & \varepsilon_{1}(\lambda)|0\rangle =
W^-_1(\lambda)
w_1^{2L}(\lambda)|0\rangle,\label{transformB}\\
\hat{A}_{aa}(\lambda)|0\rangle & = &
\varepsilon_{a+1}(\lambda)|0\rangle =
W^{-}_{a+1}(\lambda)w_3^{2L}(\lambda)|0\rangle,\\
\hat{D}(\lambda)|0\rangle & = & \varepsilon_{4}(\lambda)|0\rangle =
w^{-}_{4}(\lambda )w_7^{2L}(\lambda)|0\rangle.
\label{transform3}
\end{eqnarray}
In Appendix \ref{sec-commutation}, we give some useful relations for the
eigenvalue problem (\ref{fact}), the factorization of the transfer
matrix (\ref{TM2}) acting on the pseudo-vacuum state
(\ref{pseudo-vacuum}). We note that the operators
$\tilde{B}_a(\lambda)$, $a=1,2$, constitute a two-component vector with
both positive and negative pseudo-momenta still playing the role of the
creation fields acting on the pseudo-vacuum state. In Eq.\ (\ref{UM}),
$\tilde{E}_a(\lambda)$ are the components of the dual creation fields,
whereas one can show that the operators $\tilde{C}_i(\lambda)$ are still
the annihilation fields acting on the pseudo-vacuum state. The
integrability of the model leads to a perfect factorization of the
transfer matrix acting not only on the pseudo-vacuum state but also on
the multi-particle states. Conversely, the factorization of the transfer
matrix on the pseudo-vacuum state reveals the consistency between the
boundary reflection $K_{\pm}$-matrices and the integrability of the
model.

In order to make further progress we return to the graded RE
(\ref{RE1}) and derive commutation relations between the diagonal
fields and the creation fields. In general, the algebraic Bethe-ansatz
solution for an integrable model with open BC is more complicated than
that in the case of periodic BC due to the appearance of positive and
negative rapidities.  As in the case of the open Hubbard and Bariev
chains \cite{G1}, we find --- substituting (\ref{UM}) into (\ref{RE1})
--- that the eigenvectors of the transfer matrices are generated only
by two classes of creation fields. The first class consists of the
non-commuting vectors $\tilde{B}_{a}(\lambda)$ satisfying the
commutation relations
\begin{eqnarray}
\lefteqn{\tilde{B}_{a}(\lambda)\otimes\tilde{B}_{b}(\nu)}\nonumber\\
& = &
\frac{1}{w_1(\lambda-\nu)}\left[\tilde{B}_{c}(\nu)\otimes\tilde{B}_{d}(\lambda)
+\frac{w_{10}(\lambda+\nu)}{w_3(\lambda+\nu)}\vec{\eta }\tilde{F}(\nu)(\mathbf{I}\otimes\tilde{A}(\nu))\right]  r(\lambda-\nu) \nonumber\\
& &
-\frac{w_{10}(\lambda+\nu)}{w_3(\lambda+\nu)}\vec{\eta }\tilde{F}(\lambda)
(\mathbf{I}\otimes \tilde{A}(\nu)) +\frac{w_7(\lambda+\nu)}{w_3(\lambda+\nu)}\left[
\frac{w_{10}(\lambda-\nu)}{w_7(\lambda-\nu)}\tilde{F}(\lambda)\tilde{B}(\nu)\right.
\nonumber\\
& &
\left.-\frac{w_6(\lambda-\nu)w_7(\lambda-\nu)+w_{10}(\lambda-\nu)w_5(\lambda-\nu)}
{w_1(\lambda-\nu)w_7(\lambda-\nu)}\tilde{F}(\nu)\tilde{B}(\lambda)
\right]\vec{\eta },\label{com1}
\end{eqnarray}
where the functions $w_{i}(\mu)$ are those defined in (\ref{eq:w})
with $x=\e^{\lambda}$. The second class of creation fields contains
$\tilde{F}(\lambda)$, of (\ref{TIM}), which commute among themselves, i.e.,
$\left[\tilde{F}(\lambda), \tilde{F}(\nu)\right]=0$.  Here,
$\vec{\eta}$ is a vector defined by $\vec{\eta }=(0,1,1,0)$, $\mathbf{I}$ is a
$2\times 2$ identity matrix, and $\tilde{A}$ denotes a submatrix
given as
\begin{equation}
\tilde{A}(\nu)  =  \left( \begin {array} {cc}
\tilde{A}_{11}(\nu)&\tilde{A}_{12}(\nu)\\
\tilde{A}_{21}(\nu)&\tilde{A}_{22}(\nu)
\end {array} \right).
\end{equation}
Performing
the standard procedure, which is to keep the diagonal fields always on the
right-hand sides in the commutation relations, and after several steps of
substitutions, we arrive at the following  commutation relations
\begin{eqnarray}
\tilde{B}(\lambda)\tilde{B}_a(\nu) & = &
 \frac{w_1(\nu-\lambda)w_3(\lambda+\nu)}{w_3(\nu-\lambda)w_1(\lambda+\nu)}
 \tilde{B}_a(\nu)\tilde{B}(\lambda) + {\rm u.t.},\label{com2}\\[1ex]
\hat{D}(\lambda)\tilde{B}_a(\nu) & = &
 \frac{w_3(\lambda-\nu)
  \left[w_3^2(\lambda+\nu)-w_6(\lambda+\nu)w_{10}(\lambda+\nu)\right]}
 {w_7(\lambda-\nu)w_7(\lambda+\nu)w_3(\lambda+\nu)}
 \tilde{B}_a(\nu)\hat{D}(\lambda)\nonumber\\
& &  + {\rm u.t.},  \label{com3}\\[1ex]
\hat{A}_{ab}(\lambda)\tilde{B}_c(\nu) & = &
 \frac{w_1(\nu+\lambda)-w_2(\lambda+\nu)w_4(\lambda+\nu)}
 {w_3(\nu-\lambda)w_3(\nu+\lambda)w_1(\lambda+\nu)}
 \sum_{d,e,f,g=1}^{2}\left\{r(\lambda+\nu-2\mathrm{i}\gamma)^{ea}_{gf}
\right.\nonumber\\
& & \left. r(\lambda-\nu)^{df}_{cb}\tilde{B}_e(\nu)\hat{A}_{gd}(\lambda)\right\} +
 \rm{u.t.}\quad.\label{com4}
\end{eqnarray}
Here $a,b,c=1,2$. In the commutation relations
(\ref{com2})--(\ref{com4}), we omit all unwanted terms (u.t.) because
they consist of a complex mixture of creation and annihilation fields
and need a lot of space to display.  The complexity of the unwanted
terms plus the appearance of negative pseudo-momenta makes it very
hard to perform the algebraic Bethe ansatz in a systematic way, in
contrast to the case of the 1D Hubbard model with periodic BC
\cite{Mar}.  However, we notice that the first term in each of the
commutation relations (\ref{com2})--(\ref{com4}) contribute to the
eigenvalues of the transfer matrix which should be analytic functions
of the spectral parameter $\lambda $.  Consequently, the residues at
singular points must vanish. This yields the Bethe-ansatz equations
which in turn assure the cancellation of the unwanted terms in the
eigenvalues of the transfer matrix. Hence we prefer to use the
analytical properties rather than an analysis of the unwanted terms to
derive the Bethe-ansatz equations.  Fortunately, Eq.~(\ref{com4})
reveals a hidden SU$(1)$-symmetry structure of the nesting transfer matrix,
which is realized by the auxiliary $r$-matrix given by
\begin{equation}
r(\lambda) =
\left(\matrix{1&0&0&0\cr 0&\varrho_1(\lambda)&\varrho_2(\lambda)&0\cr
0&\varrho_2(\lambda)&\varrho_1(\lambda)&0 \cr 0&0&0&1\cr }\right)
\label{r12}
\end{equation}
with
\begin{equation}
\varrho_1(\lambda)  =
w_8(\lambda)-\frac{w_6(\lambda)w_{10}(\lambda)}{w_7(\lambda)},\quad
\varrho_2(\lambda)  =
w_9(\lambda)-\frac{w_6(\lambda)w_{10}(\lambda)}{w_7(\lambda)}.
\end{equation}
This hidden symmetry, leading to a factorization of the spin sector,
plays a crucial role in the exact solution of the model (\ref{Ham}).
For $q=\e^{\mathrm{i}\gamma}$, $0\leq \gamma\leq \pi$, the matrix
(\ref{r12}) is nothing but the scattering matrix of the anisotropic
Heisenberg XXZ model with Boltzmann weights
\begin{equation}
\varrho_1(\lambda)=\frac{\sinh 2\mathrm{i}\gamma}
                  {\sinh(\lambda+2\mathrm{i}\gamma)},\quad
\varrho_2(\lambda)=\frac{\sinh \lambda}{\sinh(\lambda+2\mathrm{i}\gamma)}.
\end{equation}
Thus, the deformation parameter $\gamma$ also plays the role of an
anisotropy parameter in the hidden XXZ open chain.  We also
note that the commutation relation (\ref{com1}) exhibits an
important symmetry, i.e.,
\begin{equation}
\vec{\eta }\, r(\lambda)=
\frac{\left[w_6(\lambda)w_7(\lambda)+
w_{10}(\lambda)w_5(\lambda)\right]
w_7(-\lambda)}{w_7(\lambda)w_{10}(-\lambda)}\vec{\eta },
\end{equation}
leading to a symmetrization of the multi-particle states.  

Following the argument of Refs.~\cite{Mar,G1}, we phenomenologically
construct the $n$-particle state,
\begin{equation}
|\Phi_n(\nu_1,\ldots,\nu_n)\rangle=
\Phi_{n}(\nu_1,\ldots,\nu_n)F^{a_1,\ldots ,a_n}|0\rangle .
\label{np-state}
\end{equation}
Here $F^{\mbox{\scriptsize $a_1,\ldots ,a_n$}}$ are the coefficients
of arbitrary linear combinations of the vectors reflecting the `spin'
degrees of freedom with $a_i=1, 2$ ($i=1, \ldots, n$).
The $n$-particle vector $\Phi_n$ satisfies the recursion
\begin{eqnarray}
\Phi_{n}(\nu_1,\ldots,\nu_n)
&=& \tilde{B}_{e_1}(\nu_1)\otimes\Phi_{n-1}(\nu_2,\ldots,\nu_n)  \label{np-vector} \\
& & -\sum _{j=2}^{n}\left\{\left[\vec{\eta }\otimes \tilde{F}(\nu_1)\right]
 \Phi_{n-2}(\nu_2,\ldots,\nu_{j-1},\nu_{j+1},\ldots,\nu_n)\right.\nonumber\\
& & \left.\left[
 \tilde{B}(\nu_j)G^{(n)}_{j-1}(\nu_1,\ldots,\nu_n)
-\left(\mathbf{I}\otimes \tilde{A}(v_j)\right)H^{(n)}_{j-1}(\nu_1,\ldots,\nu_n)
\right]\right\},\nonumber
\end{eqnarray}
where the indices $e_i=1, 2$ ($i=1, \ldots, n$) have been suppressed
on the left-hand side for brevity.  We remark that $\vec{\eta}$
excludes the possibility of two up- or two down-spin particles
residing at the same site.  $\tilde{F}$ creates a local particle pair
with opposite spins. The coefficients $G^{(n)}_{j-1}$ and
$H^{(n)}_{j-1}$ in turn can be determined from the symmetry of the
wave functions
\begin{eqnarray}
\lefteqn{\Phi_{n}(\nu_1,\ldots,\nu_j,\nu_{j+1},\ldots,\nu_n)}\nonumber\\
& = &
\frac{1}{w_1(\nu_j-\nu_{j+1})}
\Phi_{n}(\nu_1,\ldots,\nu_{j+1},\nu_{j},\ldots,\nu_n) \cdot
r(\nu_{j}-\nu_{j+1})
\end{eqnarray}
and the constraint  arising from the cancellations of the
unwanted terms in the corresponding eigenvalues of the transfer
matrix. Explicitly, the one- and  two-particle
vectors read
\begin{eqnarray}
\Phi _{1}(\nu_1) & = & \tilde{B}_{e_1}(\nu_1),\\
\Phi _{2}(\nu_1,\nu_2) & = &
 \tilde{B}_{e_1}(\nu_1)\otimes\tilde{B}_{e_2}(\nu_2)
 +\frac{w_{10}(\nu_1+\nu_2)}{w_3(\nu_1+\nu_2)}\vec{\eta}\tilde{F}(\nu_1)
 \left[\mathbf{I}\otimes\tilde{A}(\nu_2)\right] \nonumber \\
& & - \frac{w_7(\nu_1 +\nu_2)w_{10}(\nu_1 -\nu_2)}{w_3(\nu_1 +\nu_2)w_7(\nu_1 -\nu_2)}\vec{\eta}\tilde{F}(\nu_1)\tilde{B}(\nu_2),
\end{eqnarray}
respectively. Again, spin indices $e_1$ and $e_2$ are assumed
implicitly on the left-hand side such that (\ref{np-state}) is
fulfilled.

According to the algebraic Bethe ansatz, the requirement that the
unwanted terms in the eigenvalues of the transfer matrix cancel
exactly yields the so-called Bethe-ansatz equations, which are
quantization conditions for the rapidities.  The property that the
eigenvalues of the transfer matrix should have no poles suggests an
alternative way to derive the constraints on the rapidities. It turns
out that the Bethe-ansatz equations obtained by the latter way indeed
imply the integrability of the model as can be seen by checking the
consistency of the two-level transfer matrices.  Letting the diagonal
fields act on the state (\ref{np-state}) and using the commutation
relations (\ref{com2})--(\ref{com4}), we get
\begin{eqnarray}
\lefteqn{\tilde{B}(\lambda) |\Phi_n(\nu_1,\ldots,\nu_n)\rangle}\nonumber\\
& = &
 \varepsilon_{1}(\lambda)\prod_{j=1}^{n}
 \frac{\sinh\frac{1}{2}(\lambda-\nu_j+2\mathrm{i}\gamma)
       \sinh\frac{1}{2}(\lambda+\nu_j)}
      {\sinh\frac{1}{2}(\lambda-\nu_j)
       \sinh\frac{1}{2}(\lambda+\nu_j-2\mathrm{i}\gamma)}
 |\Phi _n(\nu_1,\ldots,\nu_n)\rangle + {\rm u.t.}, \label{B-eigen}\\[1ex]
\lefteqn{\hat{D}(\lambda)|\Phi _n(v_1,\ldots,v_n)\rangle}\nonumber\\
& = &
 \varepsilon_{4}(\lambda) \prod_{j=1}^{n}
 \frac{\cosh\frac{1}{2}(\lambda-\nu_j)
       \cosh\frac{1}{2}(\lambda+\nu_j-2\mathrm{i}\gamma)}
      {\cosh\frac{1}{2}(\lambda-\nu_j+2\mathrm{i}\gamma)
       \cosh\frac{1}{2}(\lambda+\nu_j)}
 |\Phi _n(\nu_1,\ldots,\nu_n)\rangle + {\rm u.t.},\label{D-eigen} \\[1ex]
\lefteqn{\hat{A}_{aa}(\lambda) |\Phi _n(\nu_1,\ldots,\nu_n)\rangle}\nonumber\\
& = &
 \varepsilon_{a+1}(\lambda)\prod_{j=1}^{n}
 \frac{\sinh\frac{1}{2}(\lambda-\nu_j+2\mathrm{i}\gamma)
       \sinh\frac{1}{2}(\lambda+\nu_j)}
      {\sinh\frac{1}{2}(\lambda-\nu_j)
       \sinh\frac{1}{2}(\lambda+\nu_j-2\mathrm{i}\gamma)}
 \nonumber\\
& & \times r(\lambda+\nu_1-2\mathrm{i}\gamma)_{g_{1}f_{1}}^{e_{1}a}
 r(\lambda-\nu_1)_{a_{1}a}^{d_{1}f_{1}}
 r(\lambda+\nu_2-2\mathrm{i}\gamma)_{g_2{}f_{2}}^{e_{2}g_{1}}
 r(\lambda-\nu_2)_{a_{2}d_{1}}^{d_{2}f_{2}}
 \nonumber \\
& & \times \ldots
 r(\lambda+\nu_n-2\mathrm{i}\gamma)_{g_{n}f_{n}}^{e_{n}g_{n-1}}
 r(\lambda-\nu_n)_{a_{n}d_{n-1}}^{d_{n}f_{n}}
 |\Phi_n(\nu_1,\ldots,\nu_n)\rangle + {\rm u.t.}
 \label{hiddenTM}
\end{eqnarray}
In the last equation above, the summation convention is implied for
the repeated indices $d_j,f_j$ and $g_j$ except for the indices
$a$ and $e_j$.  The $n$-particle state with indices $e_j$ on the
right-hand side, which is as defined in (\ref{np-state}), should be
equivalent to the one with indices $a_j$ on the left-hand side.  It is
easily found that (\ref{hiddenTM}), the eigenvalue of the submatrix
$\hat{A}_{aa}(\lambda)$, involves a nesting double-row transfer matrix
consisting of the inhomogeneous Lax operators
$r(\lambda+\nu_{n}-2\mathrm{i}\gamma)$ and $r(\lambda-\nu_n)$.  For
convenience, we shift the rapidities, $\lambda=u+\mathrm{i}\gamma$ and
$\nu_j=v_j+\mathrm{i}\gamma$, and obtain
\begin{eqnarray}
\lefteqn{\tau(u)|\Phi _n(v_1,\ldots,v_n)\rangle = \Lambda(u,\left\{v_j\right\} )|\Phi _n(v_1,\ldots,v_n)\rangle }\nonumber\\
& = &
 \left\{W_{1}^{+}(u+\mathrm{i}\gamma)\varepsilon _{1}(u+\mathrm{i}\gamma)
 \prod_{j=1}^{n}
 \frac{\sinh\frac{1}{2}(u-v_{j}+2\mathrm{i}\gamma)
       \sinh\frac{1}{2}(u+v_{j}+2\mathrm{i}\gamma)}
      {\sinh\frac{1}{2}(u-v_{j})\sinh\frac{1}{2}(u+v_{j})}\right.
\nonumber \\
& &
 \left. + W^{+}_{2}(u+\mathrm{i}\gamma)\varepsilon _{a+1}(u+\mathrm{i}\gamma)
 \right.\nonumber\\
& &
 \left.
 \times
 \prod_{j=1}^{n}
 \frac{\sinh\frac{1}{2}(u-v_{j}+2\mathrm{i}\gamma)
       \sinh\frac{1}{2}(u+v_{j}+2\mathrm{i}\gamma)}
      {\sinh\frac{1}{2}(u-v_{j})\sinh\frac{1}{2}(u+v_{j})}
 \Lambda^{(1)}(u,\{v_j\})\right.\nonumber \\
& &
 \left. + W^{+}_{4}(u+\mathrm{i}\gamma)\varepsilon _{4}(u+\mathrm{i}\gamma)
 \prod_{j=1}^{n}
 \frac{\cosh\frac{1}{2}(u-v_{j})\cosh\frac{1}{2}(u+v_{j})}
      {\cosh\frac{1}{2}(u-v_{j}+2\mathrm{i}\gamma)
       \cosh\frac{1}{2}(u+v_{j}+2\mathrm{i}\gamma)}\right\} \nonumber\\
& & \times
 |\Phi_n(v_1,\ldots,v_n)\rangle \label{T-eigen}
\end{eqnarray}
provided that
\begin{equation}
-\left.\frac{W^{+}_{1}(u+\mathrm{i}\gamma)
       W^{-}_{1}(u+\mathrm{i}\gamma)
       w_1^{2L}(u+\mathrm{i}\gamma)}
      {W^{+}_{2}(u+\mathrm{i}\gamma)
       W^{-}_{2}(u+\mathrm{i}\gamma)
       w_3^{2L}(u+\mathrm{i}\gamma)}\right|_{u=v_j}
 = \Lambda^{(1)}(u=v_j,\{v_j\}),
\end{equation}
for $j=1, \ldots, n$.  Here $\Lambda^{(1)}(u,\{v_j\})$ are  the  eigenvalues
of the nesting transfer matrix (\ref{NTM}):
\begin{equation}
\tau^{(1)}(u,\{v_j\})F^{e_{1},\ldots,e_{n}}=
\Lambda^{(1)}(u,\{v_j\})F^{e_{1},\ldots,e_{n}}.
\label{nev}
\end{equation}
The nesting transfer matrix reads
\begin{equation}
\tau^{(1)}(u,\{v_j\})=
{\rm Tr}_{0}\left[T^{(1)}(u)\bar{T}^{(1)}(u)\right],
\label{NTM}
\end{equation}
where
\begin{eqnarray}
T^{(1)}(u) & = &
 r_{12}(u+v_1)_{f_{1}g_{1}}^{e_{1}a}\ldots
 r_{12}(u+v_n)_{f_{n}g_{n}}^{e_{n}g_{n-1}}, \\
\bar{T}^{(1)}(u) &= &{T^{(1)}}^{-1}(-u) =
 r_{21}(u-v_n)_{d_{n-1}a_{n}}^{d_{n}f_{n}}\ldots
 r_{21}(u-v_1)_{aa_{1}}^{d_{1}f_{1}}.
\label{Tinv}
\end{eqnarray}
In the previous two expressions, we used the standard notation
$r_{12}(u)=p\cdot r(u)$.
Here $p$ is a standard permutation operator, which can be represented by
a $2^2\times 2^2$ matrix, i.e., $ p_{\alpha\beta, \gamma
\delta}=\delta_{\alpha\delta}\delta_{\beta\gamma}$.
We also  note that $r_{12}=r_{21}$ for the $r$-matrix (\ref{r12}) and
the trace operation in
(\ref{NTM}) leads to the identification $g_n=d_n=a$.
It can be seen that the coefficients $F^{e_{1},\ldots,e_{n}}$
act as  the multi-particle vectors for the inhomogeneous transfer
matrix (\ref{NTM}), which characterizes the spin sector of the model.

So far, we managed to solve the charge degrees of freedom. The next
task, the diagonalization of the anisotropic Heisenberg XXZ model with
open boundaries, was done previously \cite{EK}. Thus we immediately
obtain the eigenvalue of the nested transfer matrix (\ref{NTM}) as
\begin{eqnarray}
\lefteqn{\Lambda^{(1)}(u,\{u_1,\ldots,u_M\},\{v_1,\ldots,v_n\})}\nonumber\\
& = &
 \frac{2\sinh(u+2\mathrm{i}\gamma)\cosh u}{\sinh2(u+\mathrm{i}\gamma)}
 \prod_{l=1}^{M}
 \frac{\sinh(u-u_l-2\mathrm{i}\gamma)\sinh(u+u_l)}
      {\sinh(u-u_l)\sinh(u+u_l+2\mathrm{i}\gamma)}\nonumber \\
& & +
 \frac{2\cosh(u+2\mathrm{i}\gamma)\sinh u}{\sinh2(u+\mathrm{i}\gamma)}
 \prod^{n}_{j=1}
 \frac{\sinh(u-v_j)\sinh(u+v_j)}
      {\sinh(u-v_j+2\mathrm{i}\gamma)\sinh(u+v_j+2\mathrm{i}\gamma)}
 \nonumber\\
& & \times
 \prod^{M}_{l=1}
 \frac{\sinh(u-u_l+2\mathrm{i}\gamma)\sinh(u+u_l+4\mathrm{i}\gamma)}
      {\sinh(u-u_l)\sinh(u+u_l+2\mathrm{i}\gamma)}
 \label{nev2}
\end{eqnarray}
provided that
\begin{eqnarray}
\lefteqn{
 \frac{\cosh^2(u_k+2\mathrm{i}\gamma)}
     {\cosh^2 u_k}
 \prod_{j=1}^{n}
 \frac{\sinh(u_k-v_j)\sinh(u_k+v_j)}
      {\sinh(u_k-v_j+2\mathrm{i}\gamma)
       \sinh(u_k+v_j+2\mathrm{i}\gamma)}}\nonumber\\
& = & \prod_{\stackrel{\scriptstyle l=1}{l\neq k}}^{M}
 \frac{\sinh(u_k-u_l-2\mathrm{i}\gamma)\sinh(u_k+u_l)}
      {\sinh(u_k-u_l+2\mathrm{i}\gamma)\sinh(u_k+u_l+4\mathrm{i}\gamma)},\quad
\label{nbe}
k=1, \ldots, M,
\end{eqnarray}
where $M$ is the number of itinerant electrons with spin down, and $n$
is the total number of itinerant electrons. The eigenvalue
(\ref{nev2}) and the constraint (\ref{nbe}) on the spin rapidities
$u_k$ and $u_l$ have paved the way to diagonalize the transfer matrix
(\ref{TM2}) completely. Making a further shift of the spin rapidities,
i.e., $u_k=\mu_k -\mathrm{i}\gamma$ and $u_l=\mu_l -\mathrm{i}\gamma$,
the eigenvalues of the transfer matrix
(\ref{T-eigen})
\begin{equation}
\tau(u) |\Phi_n(v_1,\ldots,v_n)\rangle =
\Lambda(u,\{\mu_l\},\{v_j\}) |\Phi_n(v_1,\ldots,v_n)\rangle
\end{equation}
are given by
\begin{eqnarray}
\Lambda(u,\{\mu_l\},\{v_j\})
& = &
 \prod_{j=1}^{n}
 \frac{\sinh\frac{1}{2}(u-v_{j}+2\mathrm{i}\gamma)
       \sinh\frac{1}{2}(u+v_{j}+2\mathrm{i}\gamma)}
      {\sinh\frac{1}{2}(u-v_{j})\sinh\frac{1}{2}(u+v_{j})}\nonumber\\
& & \times
 \left\{
 W_{1}^{+}(u+\mathrm{i}\gamma)
 W_{1}^{-}(u+\mathrm{i}\gamma)
 w_1^{2L}(u+\mathrm{i}\gamma)\vphantom{\prod_{l=1}^{n}}\right.\nonumber\\
& & \left. +
 W_{2}^{+}(u+\mathrm{i}\gamma)
 W_{2}^{-}(u+\mathrm{i}\gamma)
 w_3^{2L}(u+\mathrm{i}\gamma) \right.\nonumber\\
& & \left. \times
 \prod_{l=1}^{M}
 \frac{\sinh(u-\mu_{l}-\mathrm{i}\gamma)
       \sinh(u+\mu_{l}-\mathrm{i}\gamma)}
      {\sinh(u-\mu_{l}+\mathrm{i}\gamma)
       \sinh(u+\mu_{l}+\mathrm{i}\gamma)}\right\}\nonumber\\
& & +
 \prod_{j=1}^{n}
 \frac{\cosh\frac{1}{2}(u-v_{j})\cosh\frac{1}{2}(u+v_{j})}
      {\cosh\frac{1}{2}(u-v_{j}+2\mathrm{i}\gamma)
       \cosh\frac{1}{2}(u+v_{j}+2\mathrm{i}\gamma)}\nonumber\\
& & \times
 \left\{
 W_{4}^{+}(u+\mathrm{i}\gamma)
 W_{4}^{-}(u+\mathrm{i}\gamma)
 w_7^{2L}(u+\mathrm{i}\gamma)\vphantom{\prod_{l=1}^{n}}\right.\nonumber\\
& & \left. +
 W_{2}^{+}(u+\mathrm{i}\gamma)
 W_{2}^{-}(u+\mathrm{i}\gamma)
 w_3^{2L}(u+\mathrm{i}\gamma)\right.\nonumber\\
& & \left. \times
 \prod_{l=1}^{M}
 \frac{\sinh(u-\mu_{l}+3\mathrm{i}\gamma)
       \sinh(u+\mu_{l}+3\mathrm{i}\gamma)}
      {\sinh(u-\mu_{l}+\mathrm{i}\gamma)
       \sinh(u+\mu_{l}+\mathrm{i}\gamma)}\right\}.\label{eigenv}
\end{eqnarray}
The rapidities of the charge and spin degrees of freedom satisfy the
following Bethe-ansatz equations
\begin{eqnarray}
\lefteqn{
 \zeta(v_{j},\xi^{+}) \zeta(v_{j},\xi^{-})
 \frac{\sinh^{2L}\frac{1}{2}(v_{j}-\mathrm{i}\gamma)}
      {\sinh^{2L}\frac{1}{2}(v_{j}+\mathrm{i}\gamma)}}\nonumber\\
& = &
 \prod^{M}_{l=1}
 \frac{\sinh(v_{j}-\mu_{l}-\mathrm{i}\gamma)
       \sinh(v_{j}+\mu_{l}-\mathrm{i}\gamma)}
      {\sinh(v_{j}-\mu_{l}+\mathrm{i}\gamma)
       \sinh(v_{j}+\mu_{l}+\mathrm{i}\gamma)},
\label{Bethe1}\\[1ex]
\lefteqn{
 \frac{\cosh^{2}(\mu_{k}+\mathrm{i}\gamma)}
      {\cosh^{2}(\mu_{k}-\mathrm{i}\gamma)}
 \prod_{j=1}^{n}
 \frac{\sinh(\mu_{k}-v_{j}-\mathrm{i}\gamma)
       \sinh(\mu_{k}+v_{j}-\mathrm{i}\gamma)}
      {\sinh(\mu_{k}-v_{j}+\mathrm{i}\gamma)
       \sinh(\mu_{k}+v_{j}+\mathrm{i}\gamma)}}\nonumber\\
& = &
 \prod_{\stackrel{\scriptstyle l=1}{l\neq k}}^{M}
 \frac{\sinh(\mu_{k}-\mu_{l}-2\mathrm{i}\gamma)
       \sinh(\mu_{k}+\mu_{l}-2\mathrm{i}\gamma)}
      {\sinh(\mu_{k}-\mu_{l}+2\mathrm{i}\gamma)
       \sinh(\mu_{k}+\mu_{l}+2\mathrm{i}\gamma)},
\label{Bethe2}
\end{eqnarray}
for $j=1,\ldots ,n$ and  $k=1,\ldots ,M$, respectively. Here, we introduced
the  notation
\begin{equation}
\zeta(v_{j},\xi^{\pm}) =
 \frac{\cosh\frac{1}{2}(v_j-\xi^{\pm })}
      {\cosh\frac{1}{2}(v_j+\xi^{\pm })}.
\end{equation}
{}From (\ref{HTR}) and  (\ref{eigenv}),   we obtain the
eigenvalues of the Hamiltonian (\ref{Ham}) as
\begin{eqnarray}
E & = &
 \sum_{j=1}^{n}
 \frac{2\sin^2\gamma}{\cos \gamma-\cosh v_j}+
 2L\cos\gamma-
 \frac{2}{\cos\gamma} -\sin\gamma\left[
 \cot\frac{1}{2}(\gamma-\xi^{+})\right.\nonumber\\
& &
\left.+ \tan\frac{1}{2}(\gamma+\xi^{+})+\cot\frac{1}{2}(\gamma-\xi^{-})+
 \tan\frac{1}{2}(\gamma+\xi^{-})\right].
\label{energy}
\end{eqnarray}
Apart from a shift of the boundary parameters $\xi^{\pm}$, the
Bethe-ansatz equations (\ref{Bethe1}) and (\ref{Bethe2}) coincide with
those obtained by the coordinate Bethe ansatz in Ref.~\cite{MG1}. Notice
that, besides the obtained Bethe-ansatz equations, we in addition
presented a systematic way to formulate the algebraic Bethe ansatz for
Hubbard-like models with open boundary fields and obtained the
eigenvalue of the transfer matrix (\ref{eigenv}), which is essential for
the investigation of finite-temperature properties of the model
\cite{thermo}. Furthermore, the two-level transfer matrices,
characterizing the charge and spin sectors separately, allow us to embed
different impurities into the system.  From the Bethe solutions
(\ref{Bethe1}), it is found that the boundary fields characterized by
$\zeta(v_j,\xi^{\pm})$ act indeed nontrivially on the densities of roots
for spin rapidity $\mu _l$ and charge rapidity $v_j$, and thus change
the ground state properties as well as the low-lying energy spectrum.
The function $\zeta(v_j,\xi^{\pm})$ contributes a phase shift to the
density of roots of the rapidities.  Though the first factor on the
left-hand side of the Bethe-ansatz equation (\ref{Bethe2}) originates
from the pure boundary effect of the spin sector, it contributes to both
charge and spin rapidities in a nontrivial way. Of course, we may treat
other boundary conditions for the model in a similar way. In the
following section, we are going to discuss the boundary and impurity
effects on the ground-state properties of the model at half filling.

\section{The ground-state properties}
\label{sec4}

Due to the boundary effects, the Bethe-ansatz equations (\ref{Bethe1})
and (\ref{Bethe2}) are not merely a doubling of the Bethe equations
for the periodic $\mbox{U}_q[\mbox{osp}(2|2)]$ chain \cite{doub}. Thus
the boundary fields contribute nontrivially to the ground-state
properties of the model. If we consider the ground state at half
filling, corresponding to the case that the $v_j$ are real, while the
spin variables form strings of type $\mu +\mathrm{i}\frac{\pi}{2}$,
the discrete Bethe-ansatz equations (\ref{Bethe1}) and (\ref{Bethe2})
may be written as
\begin{eqnarray}
\lefteqn{\textstyle
2L\theta_{1}(\frac{1}{2}v_{j},\frac{1}{2}\gamma)-
\theta_{2}(\frac{1}{2}v_{j},\frac{1}{2}\xi^{+})-
\theta_{2}(\frac{1}{2}v_{j},\frac{1}{2}\xi^{-})}\nonumber\\
\mbox{ } & = &
 2\pi I_{j} -
 \sum_{l=1}^{M}\left[\theta_{2}(v_{j}-\mu_{l},\gamma)+
                     \theta_{2}(v_{j}+\mu_{l},\gamma)\right],\\
\lefteqn{
 2\theta_{1}(\mu_{k},\gamma)} \nonumber\\
& = &
 2\pi J_{k} -\sum_{\stackrel{\scriptstyle l=1}{l\neq k}}^{M}
 \left[\theta_{1}(\mu_{k}-\mu_{l},2\gamma)+
       \theta_{1}(\mu_{k}+\mu_{l},2\gamma)\right]
 \nonumber\\
& & \mbox{ }
 -\sum_{j=1}^{n}\left[\theta_{2}(\mu_{k}-v_{j},\gamma)+
                     \theta_{2}(\mu_{k}+v_{j},\gamma)\right],
\end{eqnarray}
for $j=1, \ldots, n, $ and $k=1, \ldots, M$, respectively. Here we
introduced the  shift functions
\begin{eqnarray}
\theta_{1}(v,\gamma) & = &
 -\mathrm{i}\ln
 \frac{\sinh(\mathrm{i}\gamma-v)}{\sinh(\mathrm{i}\gamma+v)}
 \equiv
 2\,\mathrm{arctan}\left(\tanh v\cot\gamma\right),\\
\theta_{2}(v,\gamma) & = &
 -\mathrm{i}\ln
 \frac{\cosh(\mathrm{i}\gamma+v)}{\cosh(\mathrm{i}\gamma-v)}
 \equiv
 2\,\mathrm{arctan}\left(\tanh v\tan\gamma\right).
\end{eqnarray}
The integers $I_j$ and $J_k$ may be regarded as quantum numbers
associated to the Bethe-ansatz equations.  If we define $I_{-j}=-I_j$,
$J_{-k}=-J_k$, $v_{-j}=-v_j$, and $\mu_{-l}=-\mu_{l}$, and pass to the
thermodynamic limit $L\rightarrow\infty$, $n\rightarrow\infty$, and
$M\rightarrow\infty$, with $n/L$ and $M/L$ kept finite, the
thermodynamic Bethe-ansatz equations read
\begin{eqnarray}
\rho^{\rm c}_{\infty }(v) & = &
 \frac{1}{\pi}\left\{
 \frac{\d}{\d v}\theta_{1}({\textstyle\frac{1}{2}v,\frac{1}{2}\gamma})+
 \frac{1}{2L}\frac{\d}{\d v}\theta_{{\rm cb}}(v,\xi^{\pm})
 \right.\nonumber\\
& & \left.
 +\frac{1}{2}\int_{-\infty}^{\infty}\d\mu\frac{\d}{\d v}
 \theta_{2}(v-\mu,\gamma)\rho^{\rm s}_{\infty}(\mu)\right\},
 \label{th-bethe1}\\
\rho^{\rm s}_{\infty}(\mu) & = &
 \frac{1}{2\pi}\left\{\frac{1}{L}\frac{\d}{\d\mu}
 \theta_{{\rm sb}}(\mu,\gamma)+
 \int_{-\infty}^{\infty }\d v\frac{\d}{\d\mu}\theta_{2}(\mu-v,\gamma)
 \rho^{\rm c}_{\infty}(v)\right.\nonumber\\
& &
 \left.+\int_{-\infty}^{\infty }\d v\frac{\d}{\d\mu}
 \theta_{1}(\mu-v,2\gamma)\rho^{\rm s}_{\infty}(v)\right\}.
 \label{th-bethe2}
\end{eqnarray}
Here $\rho^{\rm c}_{\infty}(v)$ and $\rho^{\rm s}_{\infty}(\mu)$ denote
the densities of roots of the charge and spin rapidities at half filling,
respectively. The term
\begin{equation}
\theta_{{\rm cb}}(v,\xi^{\pm}) =
-\theta_{2}(v,\gamma)
- \theta_{2}({\textstyle\frac{1}{2}v,\frac{1}{2}\xi^{+}}) -
 \theta_{2}({\textstyle\frac{1}{2}v,\frac{1}{2}\xi^{-}})
\end{equation}
characterizes the charge contributions to the densities of roots from
the boundary potentials, whereas
\begin{equation}
\theta_{{\rm sb}}(\mu,\gamma) =
 2\theta_{1}(\mu,\gamma)-\theta_{1}(\mu,2\gamma)-
 \theta_{1}(2\mu,2\gamma)-\theta_{2}(\mu,\gamma)
\end{equation}
denotes the boundary contributions in the spin sector.  We would like
to stress that, although $\theta_{{\rm cb}}(v,\xi^{\pm })$ arises
completely from the charge degrees of freedom, and $\theta_{{\rm
sb}}(\mu,\gamma)$ only from the spin degrees of freedom, both terms
contribute nontrivially to the densities of roots $\rho^{\rm
c}_{\infty}(v)$ and $\rho^{\rm s}_{\infty}(\mu)$. We also find that
the ground state of the system is a singlet, which means that the
variables $\mu$ and $v$ occupy the entire interval from $-\infty$ to
$\infty$.  By Fourier transformation, the solutions of
(\ref{th-bethe1}) and (\ref{th-bethe2}) have the form
\begin{eqnarray}
\rho^{\rm c}_{\infty}(v) & = &
 \frac{1}{2\pi}\int_{-\infty}^{\infty}
 \hat{\rho}^{\rm c}_{\infty}(\omega)\e^{-\mathrm{i}\omega v}\d\omega,
\quad
 \hat{\rho}^{\rm c}_{\infty}(\omega)=\hat{\rho}^{\rm c}_{0}(\omega)+
 \frac{1}{L}\hat{\rho}^{\rm c}_{{\rm b}}(\omega),\label{density1}
\\
\rho^{\rm s}_{\infty}(\mu) & = &
 \frac{1}{2\pi}\int_{-\infty}^{\infty}
 \hat{\rho}^{\rm s}_{\infty}(\omega)\e^{-\mathrm{i}\omega\mu}\d\omega,\quad
 \hat{\rho}^{\rm s}_{\infty}(\omega)=\hat{\rho}^{\rm s}_{0}(\omega)+
 \frac{1}{L}\hat{\rho}^{\rm s}_{{\rm b}}(\omega).\label{density2}
\end{eqnarray}
Here, $\hat{\rho}^{\rm c}_{0}(\omega)$ and
$\hat{\rho}^{\rm s}_{0}(\omega)$ denote the bulk densities of roots for
the charge and the spin rapidities, which are given by
\begin{equation}
\hat{\rho}^{\rm c}_{0}(\omega) =
 \frac{2\sinh(\frac{\pi}{2}-\gamma)\omega}
      {\cosh\frac{\pi}{2}\omega},\,\,\,
\hat{\rho}^{\rm s}_{0}(\omega)  =
 \frac{1}{\cosh\frac{\pi}{2}\omega},
\end{equation}
respectively. The remaining parts $\hat{\rho}^{\rm c}_{{\rm b}}(\omega)$
and $\hat{\rho}^{\rm s}_{{\rm b}}(\omega )$ contain
the contributions caused by the boundary terms, i.e.,
\begin{eqnarray}
\hat{\rho}^{\rm c}_{{\rm b}}(\omega) & = & -
\frac{\left[\sinh\xi^{+}\omega +\sinh\xi^{-}\omega\right]
\cosh(\frac{\pi }{2}-\gamma)\omega}
{\sinh(\pi-\gamma)\omega\,\cosh\frac{\pi}{2}\omega}
-\frac{\sinh\frac{\gamma}{2}\omega}
{\sinh(\frac{\pi}{2}-\frac{\gamma}{2})\omega}, \label{bound-eff-c}\\
\hat{\rho}^{\rm s}_{{\rm b}}(\omega) & = &
-\frac{\left[\sinh\xi^{+}\omega +\sinh\xi^{-}\omega\right]}
{2\sinh(\pi-\gamma)\omega\,\cosh\frac{\pi}{2}\omega}
+\frac{\sinh(\frac{\gamma}{2}-\gamma )\omega}
{2\sinh(\frac{\pi}{2}-\frac{\gamma}{2})\omega\,\cosh\frac{\pi}{2}\omega},
\label{bound-eff-s}
\end{eqnarray}
respectively.  In the expressions (\ref{bound-eff-c}) and
(\ref{bound-eff-s}), we separated the effects of
the boundary potentials and the pure boundary effects, i.e.,
corresponding to the first and the second terms on the right-hand sides
of (\ref{bound-eff-c}) and (\ref{bound-eff-s}), respectively. It is
straightforward to recover the result for free BC for the model
(\ref{Ham}) by switching off the boundary potentials via $\xi^{\pm
}\rightarrow 0$. Because the open BC do not spoil the symmetries
$\hat{\rho}^{\rm c}_{\infty}(\omega)=\hat{\rho}^{\rm
c}_{\infty}(-\omega)$ and $\hat{\rho}^{\rm
s}_{\infty}(\omega)=\hat{\rho}^{\rm s}_{\infty}(-\omega)$, the energy
per site of the singlet ground state $(n=\frac{L}{2})$, calculated
from (\ref{energy}), reduces to the form
\begin{eqnarray}
E & = &-4\sin\gamma\int_{0}^{\infty}
\frac{\cosh(\frac{\pi }{2}-\gamma)\omega\,\sinh(\pi-\gamma)\omega }
{\cosh\frac{\pi}{2}\omega\,\sinh\pi \omega}\d\omega+2\cos\gamma\nonumber\\
& &
-\frac{1}{L}\left\{
4\sin\gamma\int_{0}^{\infty}
 \frac{\hat{\rho}^{\rm c}_{{\rm b}}(\omega)\sinh(\pi-\gamma)\omega}
      {\sinh\pi\omega}\d\omega
 +\frac{2}{\cos\gamma}+\sin \gamma\left[\cot\frac{1}{2}(\gamma-\xi^{+})
\right.\right.\nonumber\\
& &
\left.\left.+\tan\frac{1}{2}(\gamma+\xi^{+})+
                    \cot\frac{1}{2}(\gamma-\xi^{-})
+\tan\frac{1}{2}(\gamma+\xi^{-})\right]
\vphantom{\int_{0}^{\infty}}\right\}.
\label{g-energ}
\end{eqnarray}
The first term in the ground-state energy (\ref{g-energ}) is the bulk
ground-state energy which coincides with the result of the periodic
chain \cite{doub}. We emphasize  that the boundary potentials do not
only enter in the expression for the ground state energy explicitly as
$\cot\frac{1}{2}(\gamma-\xi^{\pm})+\tan\frac{1}{2}(\gamma+\xi^{\pm})$,
but also implicitly via $\rho^{\rm c}_{\infty}(v)$ and $\rho^{\rm
s}_{\infty}(v)$ of (\ref{density1}) and (\ref{density2}).

\begin{figure}
\centerline{\psfig{file=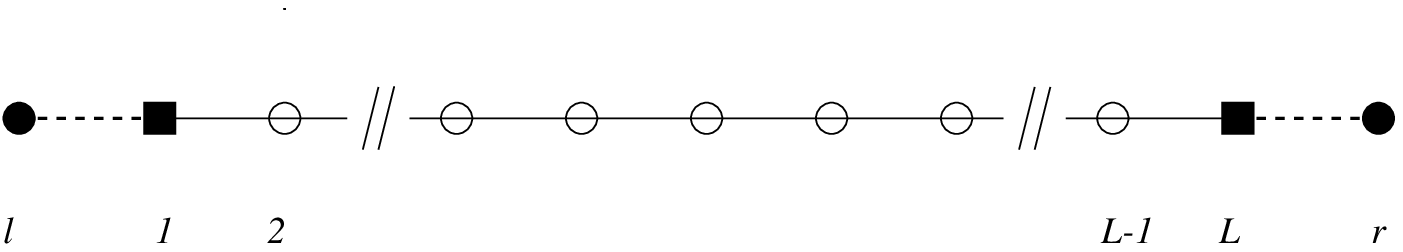,width=\figwidth}}
\caption{\label{fig-im+bmi+osp}Impurities coupled to each of the
boundaries.}
\end{figure}

Before closing this section, we discuss the problem of embedding
integrable impurities into the open chain \cite{imp2,imp3,imp4}. The
algebraic Bethe ansatz provides us with a natural way to incorporate
different kinds of impurities. If we embed two impurity vertices at the
boundaries, namely, extend (\ref{tm}) to
\begin{equation}
T(\lambda)=R_{{\rm r},0}(\lambda+p_{\rm r})
R_{L,0}(\lambda)R_{L-1,0}(\lambda)\ldots
 R_{2,0}(\lambda)R_{1,0}(\lambda)R_{\ell ,0}(\lambda+p_{\ell}),
\end{equation}
the impurity Hamiltonian with the impurity-host charge interactions
and exchange coupling between the impurities and boundaries can be
determined by
\begin{eqnarray}
H_{{\rm bi}} & = &\frac{1}{{\rm Str}_{0}K_{+}(0)}
         \left\{{\rm Str}_{0}[K_{+}(0)R^{\prime}_{{\rm r},0}(p_{\rm r})R^{-1}_{{\rm r},0}(p_{\rm r})]\right.\nonumber\\
& &\left.+ {\rm Str}_{0}[K_{+}(0)R_{{\rm r},0}(p_{{\rm r}})
         R^{\prime}_{L,0}(0)R_{L,0}^{-1}(0)R_{{\rm r},0}^{-1}(p_{{\rm r}})]
         \right\}\nonumber \\
& & + \frac{1}{2} R_{1,0}(0) R_{\ell, 0}(p_{\ell})
          K_{-}^{\prime}(0)
          R_{\ell , 0}^{-1}(p_{\ell}) R_{1,0}^{-1}(0)\nonumber\\
& &+ R_{10}(0) R_{\ell ,0}^{\prime}(p_{\ell})
          R_{\ell ,0}^{-1}(p_{\ell})R_{1,0}^{-1}(0).
\end{eqnarray}
Due to rather lengthy algebra,
we schematically present the interactions between boundaries
and impurities by Figure \ref{fig-im+bmi+osp} instead of presenting
 $H_{{\rm bi}}$ explicitly in terms of fermionic operators.
The quantities $p_{\rm r}$ and $p_{\ell}$ shifting the pseudo-momenta
characterize the impurity rapidities or impurity strength. As
discussed previously \cite{imp1,imp3},
this shift preserves the integrability of the model and allows one to
continuously vary the strength of the impurity coupling to the
boundaries revealing different impurity effects.  This embedding of
the impurities, carrying both charge and spin degrees of freedom,
leads to an additional factor
\begin{equation}
\prod_{m=\ell,{\rm r}}
 \frac{\sinh\frac{1}{2}(v_j+p_m-\mathrm{i}\gamma)
       \sinh\frac{1}{2}(v_j-p_m-\mathrm{i}\gamma)}
      {\sinh\frac{1}{2}(v_j+p_m+\mathrm{i}\gamma)
       \sinh\frac{1}{2}(v_j-p_m+\mathrm{i}\gamma)},
\end{equation}
on the left-hand side of the Bethe equation (\ref{Bethe1}), thus
contributing to the densities of roots for the charge and the spin
rapidities in (\ref{density1}) and (\ref{density2}) at order $1/L$, namely,
\begin{eqnarray}
\hat{\rho}^{\rm c}_{\infty}(\omega) & = &
 \hat{\rho}^{\rm c}_{0}(\omega)+
 \frac{1}{L}\left[\hat{\rho}^{\rm c}_{{\rm b}}(\omega)+
                  \hat{\rho}^{\rm c}_{{\rm i}}(\omega)\right],\\
\hat{\rho}^{\rm s}_{\infty}(\omega) & = &
 \hat{\rho}^{\rm s}_{0}(\omega)+
 \frac{1}{L}\left[\hat{\rho}^{\rm s}_{{\rm b}}(\omega)+
                  \hat{\rho}^{\rm s}_{{\rm i}}(\omega )\right].
\end{eqnarray}
The contributions to the densities of roots
from the impurity terms are
\begin{eqnarray}
\hat{\rho}^{\rm c}_{{\rm i}}(\omega) & = &
 \left(\cos p_{\rm r}\omega +\cos p_{\ell}\omega \right)
 \hat{\rho}^{\rm c}_{0}(\omega),\\
\hat{\rho}^{\rm s}_{{\rm i}}(\omega) & = &
 \left(\cos p_{\rm r}\omega +\cos p_{\ell}\omega \right)
 \hat{\rho}^{\rm s}_{0}(\omega).
\end{eqnarray}
In this case, the ground state energy (\ref{g-energ}) has to be
changed slightly by $\left[\hat{\rho}^{\rm c}_{{\rm
b}}(\omega)+\hat{\rho}^{\rm c}_{{\rm i}}(\omega)\right]$ replacing
$\hat{\rho}^{\rm c}_{{\rm b}}(\omega)$, which affects the surface
energy and finite-size corrections.

{}From the above discussion, one can easily distinguish the effects of
the impurities, boundary potentials, and the free edges of the system,
i.e., the effects of dynamic magnetic impurities, the external scalar
boundary fields, and open boundary conditions. In addition, the
integrability of the open chain also allows us to embed a forward
scattering impurity (without any reflection scattering amplitude) into
the bulk part.  Although the resulting impurity Hamiltonian is different
from the boundary Hamiltonian, the Bethe ansatz shows that the impurity
effects do not depend on the position of the impurity due to the pure
forward scattering that is required by integrability \cite{imp1}.
Another interesting embedding of the impurity would be given by
operator-valued boundary $K_{\pm}$-matrices \cite{imp4} which lead to
Kondo-impurity-like terms in the Hamiltonian. These different embeddings
of the impurity would provide results essential to the study of the
thermodynamic properties such as low-lying excitations, finite-size
corrections, magnetization, etc. The other string solutions to the Bethe
equations (\ref{Bethe1}) and (\ref{Bethe2}), which form the charge and
spin bound states, give an independent approach to the investigation of
the low-lying excitations for the model. We intend to consider this
situation in the future.

\section{Conclusion and discussion}
\label{sec5}

In summary, we have discussed the algebraic Bethe-ansatz solution for
the extended Hubbard model with boundary fields associated with the
quantum superalgebra $\mbox{U}_q[\mbox{osp}(2|2)]$ in terms of the
graded QISM. Two classes of solutions of the graded RE leading to four
kinds of possible boundary terms in the Hamiltonian were
obtained.  The Bethe-ansatz equations, the eigenvalue of the transfer
matrix and the energy spectrum were given explicitly.  The
ground-state properties in the thermodynamic limit were also
studied.  We found that the model exhibits an anisotropic Heisenberg
XXZ open chain as its nesting transfer matrix characterizing the spin
sector. This nesting structure seems to be different from that of
other extended Hubbard models \cite{imp4}.  The Bethe-ansatz results
allow us to embed impurities at the boundaries of the model. Our
results provide a useful starting point for studying the
thermodynamic properties and correlation functions for the model.  The
boundary potentials, pure boundary effects and impurity effects
contribute nontrivially and separately to both the density of roots of
the charge rapidity and the spin rapidity at order $1/L$. The impurity
strengths $p_{\ell}$ and $p_{r}$, the boundary parameters $\xi^{\pm}$
and the $q$-deformation parameter $\gamma$ change the asymptotic
behavior of the thermodynamic Bethe-ansatz equations, i.e., the band filling,
magnetization, susceptibility, compressibility, finite-size
corrections, etc. Our computations could open an alternative way to
study the thermodynamic properties for the integrable double
sine-Gordon model \cite{doubS} as well as to investigate the tunneling
effects in quantum wires \cite{wire}.

\begin{ack}
  X.-W.G.\ thanks M.J.\ Martins and H.-Q.\
  Zhou for helpful discussions and communications. He gratefully acknowledges
  the hospitality of the Institut f\"{u}r Physik, Technische
Universit\"{a}t Chemnitz. The work has been
  supported by the DFG via SFB393 and the Brazilian research programs
  CNPq and FAPERGS.
\end{ack}

\appendix

\section{The boundary \protect\boldmath{$K_{\pm}$}-matrices}
\label{sec-boundary-matrices}

Let us consider an algebraic ansatz which fixes the boundary
$K_{\pm}$-matrices.  Due to the isomorphism between
$K_{+ }$- and $K_{-}$-matrices, we
need to solve the RE (\ref{RE1}) to determine them. Let us first fix
the left boundary $K_-$-matrix (\ref{K-M}).
By substituting  it into the RE (\ref{RE1}),  one may find that  the
RE (\ref{RE1}) involves two variables $\lambda$ and $\nu $ which make
the functional equations involving $K_-^{(i)}(\lambda),\, i=1,\ldots,4$ much
more complicated. Nevertheless, taking into account the structure of
the $R$-matrix (\ref{R12}) and the RE (\ref{RE1}), we find that the
functional equations of the RE arising from the positions
corresponding to the permutation operator
${\bf P}_{\alpha\beta,\gamma\delta}=
(-1)^{P(\alpha)P(\beta)}\delta_{\alpha\delta}\delta_{\beta\gamma}$
provide us with simpler equations than the ones arising from other
positions in the RE. These simpler equations allow us to separate
$K_-^{(i)}(\lambda),K_-^{(i)}(\nu ) $ into factorized forms as follows
\begin{eqnarray}
\frac{K_-^{(1)}(\nu)}{K_-^{(2)}(\nu)}& = &
 \frac{w_3(\lambda+\nu)w_2(\lambda-\nu)K_-^{(1)}(\lambda)+
       w_3(\lambda-\nu)w_4(\lambda+\nu)K_-^{(2)}(\lambda)}
      {w_2(\lambda+\nu)w_3(\lambda-\nu)K_-^{(1)}(\lambda)+
       w_4(\lambda-\nu)w_3(\lambda+\nu)K_-^{(2)}(\lambda)},\label{e1}\\
\frac{K_-^{(1)}(\nu)}{K_-^{(3)}(\nu)}& = &
 \frac{w_3(\lambda+\nu)w_2(\lambda-\nu)K_-^{(1)}(\lambda)+
       w_3(\lambda-\nu)w_4(\lambda+\nu)K_-^{(3)}(\lambda)}
      {w_2(\lambda+\nu)w_3(\lambda-\nu)K_-^{(1)}(\lambda)+
       w_4(\lambda-\nu)w_3(\lambda+\nu)K_-^{(3)}(\lambda)},\label{e2}\\
\frac{K_-^{(3)}(\nu)}{K_-^{(4)}(\nu)}& = &
 \frac{w_3(\lambda+\nu)w_2(\lambda-\nu)K_-^{(3)}(\lambda)+
       w_3(\lambda-\nu)w_4(\lambda+\nu)K_-^{(4)}(\lambda)}
      {w_2(\lambda+\nu)w_3(\lambda-\nu)K_-^{(3)}(\lambda)+
       w_4(\lambda-\nu)w_3(\lambda+\nu )K_-^{(4)}(\lambda)}\label{e3}.
\end{eqnarray}
Substituting the Boltzmann weights of the $R_{12}$-matrix (\ref{R12})
into the equations above, and analyzing the structure of these
equations, we can infer that the functions on the right-hand side of
each of the equations (\ref{e1})--(\ref{e3}) involving the variable $\lambda$
should cancel because the left-hand sides of
(\ref{e1})--(\ref{e3}) are functions of the variable $\nu $ only. This
cancellation leaves us with the following expressions for $K_-^{(i)}(\lambda)$
\begin{eqnarray}
K_-^{(1)}(\lambda)&=& \left(x+C^{(1)}\right)\left(x+C^{(2)}\right)\left(x+C^{(3)}\right),\label{stru1}\\
K_-^{(2)}(\lambda)&=& \left(x^{-1}+C^{(1)}\right)\left(x+C^{(2)}\right)\left(x+C^{(3)}\right),\label{stru2}\\
K_-^{(3)}(\lambda)&=& \left(x+C^{(1)}\right)\left(x^{-1}+C^{(2)}\right)\left(x+C^{(3)}\right),\label{stru3}\\
K_-^{(4)}(\lambda)&=& \left(x+C^{(1)}\right)\left(x^{-1}+C^{(2)}\right)\left(x^{-1}+C^{(3)}\right),\label{stru4}
\end{eqnarray}
with a minimal number of coefficients that are to be determined. We
regard Eq.~(\ref{stru1})-(\ref{stru4}) as the main structure of the $K$-matrix. Here
$x=\e^{\lambda}$ as in section 2. Employing the RE again with
(\ref{stru1})--(\ref{stru4}) it is easily found that only one
coefficient is free, and we have the relations
$C^{(1)}=C^{(2)}=-C^{(3)}q^2$ or
$C^{(1)}=-C^{(3)}q^2=-q^2/C^{(2)}$. The first case yields the solution
(\ref{Km}) while the second case corresponds to (\ref{Km2}). The
corresponding boundary terms have been given by
(\ref{eq:u})--(\ref{bt2}), respectively. Due to the fact that the left
and right boundaries are independent of each other, the two cases of
boundary terms allow four possible combinations. In particular, we
find a spin-degenerate situation with chemical potentials at the
boundaries (i) identical for spin-up and spin-down electrons, of
(\ref{bt1}), or (ii) different yielding (\ref{bt2}). Or, the spins are
distinguished either (iii) at the right boundary, i.e., $V_{L,\pm}$ as
in (\ref{bt2}), or (iv) at the left boundary, i.e., $V_{1,\pm}$ as in
(\ref{bt2}), and other chemical potentials given by (\ref{bt1}).  In
all cases, $U(\xi_{\pm})$ is given by (\ref{eq:u}).  These potentials
constitute the general integrable boundary terms corresponding to the
diagonal boundary $K$-matrices. In this paper, we restrict our
discussion to the physically most realizable, symmetric case given by
(\ref{Km}).

\section{The Hamiltonian}
\label{sec-hamiltonian}

Because of ${\rm Str_0}[K_+(0)]=0$, we have to consider the
second-order expansion of the transfer matrix (\ref{TM}) with respect
to the spectral parameter $\lambda$ in order to construct the
Hamiltonian with boundary fields.  Following
\cite{open2}, we derive
\begin{eqnarray}
 H & = &
 \sum _{j=1}^{L-1}H_{j,j+1}+\frac{1}{2}{K}^{'}_-(0)
+\left\{{\rm Str}_0\left[K^{'}_{+}(0)R^{'}_{L0}(0)P_{L0}\right]
 \vphantom{\frac{1}{2}}\right. \nonumber \\
& &
 \left. +\frac{1}{2}{\rm Str}_0\left[K_+(0)R^{''}_{L0}(0)P_{L0}\right]+
\frac{1}{2}{\rm Str}_0\left[K_+(0)(R^{'}_{L0}(0)P_{L0})^2\right]
\right\}/\left\{ \vphantom{\frac{1}{2}} \right.\nonumber\\
& &
\left.
{\rm Str}_0\left[{K}^{'}_{+}(0)\right]+
2{\rm Str}_0\left[K_+(0)R^{'}_{L0}(0)P_{L0}\right]
\vphantom{\frac{1}{2}}\right\}, \label{H-bt}
\end{eqnarray}
where
\begin{equation}
\left.H_{j,j+1}=P_{j,j+1}\frac{\d\,R_{j,j+1}(\lambda)}{\d\lambda}\right| _{\lambda=0}.
\label{H-b}
\end{equation}
The prime denotes the derivative with respect to the spectral
parameter $\lambda$. Using (\ref{H-b}), we can write the bulk Hamiltonian
in terms of two commuting species of Pauli matrices
$\sigma$ and $\tau$, i.e.,
\begin{eqnarray}
H_{j,j+1} & = &\left(\sigma _j^{+}\sigma _{j+1}^{-}+
\sigma^{-}_j\sigma^{+}_{j+1}\right)
\left(\frac{\sigma_j^{z}\tau_j^{z}+
\sigma_{j+1}^{z}\tau_{j+1}^{z}}{4}-\frac{1+\tau_j^{z}\tau_{j+1}^{z}}{2(q-q^{-1})}\right)
\nonumber\\
& &
+\left(\tau_j^{+}\tau_{j+1}^{-}+\tau_j^{-}\tau_{j+1}^{+}\right)
\left(\frac{\sigma_j^{z}\tau_j^{z}+
\sigma_{j+1}^{z}\tau_{j+1}^{z}}{4}
-\frac{1+\sigma_{j}^{z}\sigma_{j+1}^{z}}{2(q-q^{-1})}\right)
\nonumber\\
& &
+\frac{q+q^{-1}}{2(q-q^{-1})}\left[
\sigma_j^{+}\sigma^{-}_{j+1}\tau_j^{-}\tau_{j+1}^{+}
-\sigma_j^{+}\sigma^{-}_{j+1}\tau_j^{+}\tau_{j+1}^{-}+ {\rm h.c.}\right]\nonumber\\
& &
-\frac{q+q^{-1}}{8(q-q^{-1})}
\left[\sigma_j^{z}\tau_{j+1}^{z}+\sigma_{j+1}^{z}\tau_{j}^{z}+
\sigma_j^{z}\tau_{j}^{z}+\sigma_{j+1}^{z}\tau_{j+1}^{z}
+4\,\mathbf{I}\right]
\nonumber\\
& &+
\frac{1}{4}\left(\sigma_j^{z}-\sigma_{j+1}^{z}+\tau_{j}^{z}-
\tau_{j+1}^{z}\right).\label{Ham-Pm}
\end{eqnarray}
For the solution  (\ref{Km}) of the RE,
the second term in (\ref{H-bt}) gives the boundary terms at site $j=1$
\begin{equation}
H_{1} = -\frac{(1+q^2)\xi_-\sigma_1^{z}\tau_1^{z}
-(2q-\xi_-+q^2\xi_-)(\sigma_1^z+\tau_1^z)}
{4(q-\xi_-)(1+q\xi_-)}.\label{bt1-1}
\end{equation}
The boundary terms at site $j=L$ are then given by the remaining terms
of (\ref{H-bt}) as
\begin{equation}
H_{L} =-\frac{(1+q^2)\xi_+\sigma_L^{z}\tau_L^{z}
-(2q\xi_++1-q^2)\xi_+(\sigma_L^z+\tau_L^z)}
{4(q-\xi_+)(1+q\xi_+)}
.\label{bt2-L}
\end{equation}
Using the Jordan-Wigner transformation \cite{J-W}, apart from a
factor $-1/(q-q^{-1})$ --- absorbed in the constant $c_2$ in
(\ref{HTR}) --- the bulk Hamiltonian (\ref{Ham-Pm}) together with
the boundary terms (\ref{bt1-1}) and (\ref{bt2-L}) has the form
presented in Eqs.~(\ref{Ham})--(\ref{bt1}).
The last term in (\ref{Ham-Pm}) should be taken into account in the boundary
terms.

\section{Useful commutation relations}
\label{sec-commutation}

For the factorization of the transfer matrix (\ref{TM}) acting on the
pseudo-vacuum state, we need the following commutation relations
\begin{eqnarray}
w_1(2\lambda)C_1(\lambda)\bar{B}_1(\lambda)& = &
 w_2(2\lambda)\left[\bar{B}(\lambda)B(\lambda)
 -A_{11}(\lambda)\bar{A}_{11}(\lambda)\right],\\
w_1(2\lambda)\bar{C}_1(\lambda)B_1(\lambda)& = &
 w_4(2\lambda)\left[B(\lambda)\bar{B}(\lambda)
 -\bar{A}_{11}(\lambda){A}_{11}(\lambda)\right],\\
w_1(2\lambda)C_2(\lambda)\bar{B}_2(\lambda)& = &
 w_2(2\lambda)\left[\bar{B}(\lambda)B(\lambda)
 -A_{22}(\lambda)\bar{A}_{22}(\lambda)\right],\\
w_1(2\lambda)\bar{C}_2(\lambda)B_2(\lambda)& = &
 w_4(2\lambda)\left[B(\lambda)\bar{B}(\lambda)
 -\bar{A}_{22}(\lambda){A}_{22}(\lambda)\right],\\
w_1(2\lambda)C_3(\lambda)\bar{F}(\lambda)& = &
 w_5(2\lambda)\left[\bar{B}(\lambda)B(\lambda)-
 D(\lambda)\bar{D}(\lambda)\right]\nonumber\\
& &
 -w_2(2\lambda)\left[{C}_4(\lambda)\bar{E}_1(\lambda)+
                   {C}_5(\lambda)\bar{E}_2(\lambda)\right],\\
w_4(2\lambda)C_3(\lambda)\bar{F}(\lambda)& = &
 w_2(2\lambda)\left[\bar{A}_{11}(\lambda)A_{11}(\lambda)-
                  {D}(\lambda)\bar{D}(\lambda)\right]
 -{C}_4(\lambda)\bar{E}_1(\lambda)\nonumber\\
& &
 +w_5(2\lambda)\bar{C}_1(\lambda)B_1(\lambda)-
 w_8(2\lambda){C}_5(\lambda)\bar{E}_2(\lambda),\\
w_4(2\lambda)C_3(\lambda)\bar{F}(\lambda)& = &
 w_2(2\lambda)\left[\bar{A}_{22}(\lambda)A_{22}(\lambda)-
 {D}(\lambda)\bar{D}(\lambda)\right]
 -{C}_5(\lambda)\bar{E}_2(\lambda)\nonumber\\
& &
 +w_5(2\lambda)\bar{C}_2(\lambda)B_2(\lambda)-
 w_8(2\lambda){C}_4(\lambda)\bar{E}_1(\lambda),
\end{eqnarray}
which can be derived directly from (\ref{YBA}). {}From these relations,
we obtain
\begin{eqnarray}
\lefteqn{C_3(\lambda)\bar{F}(\lambda) = \frac{(q^2-1)^2}
       {(x^2q^2-1)(x^2-q)} \left[\bar{A}_{11}(\lambda)A_{11}(\lambda)
+\bar{A}_{22}(\lambda)A_{22}(\lambda)\right]}\nonumber\\
& &
-\frac{2(q^2-1)}{(x^2+1)(x^2-q^2)}B(\lambda)\bar{B}(\lambda)
+\frac{2(q^2-1)}{(x^2q^2-1)(x^2+1)}
 {D}(\lambda)\bar{D}(\lambda),\label{re-1}\\[1ex]
\lefteqn{C_4(\lambda)\bar{E}_1(\lambda) =
 \frac{q^2(q^2-1)(x^4-1)}{(x^4q^4-1)(x^2-q^2)}
 \bar{A}_{11}(\lambda)A_{11}(\lambda)}\nonumber\\
& &
-\frac{x^2(q^2-1)^2(q^2+1)}{(x^4q^4-1)(x^2-q^2)}
 \bar{A}_{22}(\lambda)A_{22}(\lambda)
 -\frac{q^2-1}{x^2q^2-1}{D}(\lambda)\bar{D}(\lambda),\\[1ex]
\lefteqn{C_5(\lambda)\bar{E}_2(\lambda) =
-\frac{x^2(q^2-1)^2(q^2+1)}{(x^4q^4-1)(x^2-q^2)}
 \bar{A}_{11}(\lambda)A_{11}(\lambda)}\nonumber\\
& &
+\frac{q^2(q^2-1)(x^4-1)}{(x^4q^4-1)(x^2-q^2)}
 \bar{A}_{22}(\lambda)A_{22}(\lambda)
 -\frac{q^2-1}{x^2q^2-1}{D}(\lambda)\bar{D}(\lambda)
\label{re-3}.
\end{eqnarray}
In addition, one also can show that
$C_i(\lambda)\bar{B}_j(\lambda)=0$ for $i\neq j$, $i=1,2,3$, and
$j=1,2$. With the help of  (\ref{re-1})--(\ref{re-3}), it
is not difficult to derive (\ref{fact}).


\begin{thebibliography}{99}

\bibitem{SutR93}
B. Sutherland and R.A. {R\"{o}mer}, Phys. Rev. Lett. 71  (1993) 2789;\newline
R.A. {R\"{o}mer} and B. Sutherland, Phys. Rev. B 49 (1994) 6779.
\bibitem{nfl1}
 H. Lee and J. Toner, Phys. Rev. Lett. 69 (1992) 3378;\newline
 A. Furusaki and N. Nagaosa, Phys. Rev. Lett. 72 (1994) 892.
\bibitem{nfl2}
C.L. Kane and M.P.A. Fisher, Phys. Rev. Lett. 68 (1992) 1220;\newline
P. Fr\"ojdh and H. Johannesson, Phys. Rev. Lett. 75 (1995) 300.
\bibitem{HTc}J.B. Bednorz and K.A. M\"{u}ller, Z. Phys. B 64 (1986) 189.
\bibitem{HF}O.O. Bernal, Phys. Rev. Lett. 75 (1995) 2023; \newline
B.B. Maple, J. Low Temp. Phys. 99 (1995) 223.
\bibitem{Lutc}F.D.M. Haldane, J. Phys. C: Solid State Phys. 14 (1981) 2585.
\bibitem{Schulz}
H.J. Schulz, Int. J. Mod. Phys. B 5 (1991) 57.
\bibitem{Mattis-Lieb}D.C. Mattis, E.H. Lieb, J. Math. Phys. 6 (1965) 304.
\bibitem{Kondo}
 N. Andrei, K. Furuya and J.H. Lowenstein,
   Rev. Mod. Phys. 55 (1983) 331;\newline
 A.M. Tsvelik and P.B. Wiegmann, Adv. Phys. 32 (1983) 453;\newline
J. Stat. Phys. 38 (1985) 125.
\bibitem{QISM}
 L.D. Faddeev, in
{\it Recent advances in field theory and statistical mechanics},
   eds. J.-B. Zuber and R. Stora
   (North-Holland, Amsterdam, 1984);\newline
 P.P. Kulish and E.K. Sklyanin,
   in {\it Integrable Quantum Field Theories},
  Lecture Notes in Physics Vol.~151,
  eds. J. Hietarinta and C. Montonen
   (Springer, Berlin, 1982) p.~61;\newline
 V.E. Korepin, N.M. Bogoliubov and A.G. Izergin,
   {\it Quantum Inverse Scattering Method and Correlation Functions}
   (Cambridge University Press, Cambridge, 1993).
\bibitem{imp1}
 N. Andrei and H. Johannesson, Phys. Lett. A 100 (1984) 108;\newline
 P. Schlottmann, J. Phys. Condens. Matter 3 (1991) 6617;\newline
   J. Phys. Condens. Matter 11 (1999) 4617;
   Nucl. Phys. B 552 (1999) 727;\newline
 P. Schlottmann, Phys. Rev. B 57 (1998) 10638;\newline
 S.R. Aladim and M.J. Martins, J. Phys. A: Math. Gen. 26 (1993) L529;\newline
 A.A. Zvyagin and P. Schlottmann, Phys. Rev. B 56 (1997) 300;\newline
 H.-P. Eckle, A. Punnoose and R.A. R\"{o}mer,
    Europhys. Lett. 39 (1997) 293.
\bibitem{YBE}
 J.B. McGuire, J. Math. Phys. 5 (1964) 622;\newline
 C.N. Yang, Phys. Rev. Lett. 19 (1967) 1312;\newline
 R.J. Baxter,
 {\it Exactly Solved Models in Statistical Mechanics}
 (Academic Press, London 1982).
\bibitem{EK}
 E.K. Sklyanin, J. Phys. A: Math. Gen. 21 (1988) 2375.
\bibitem{open1}
 L. Mezincescu and R.I. Nepomechie, J. Phys. A: Math. Gen. 24 (1991) L17;
   Int. J. Mod. Phys. A8 (1991) 5231;
   Int. J. Mod. Phys. A 7 (1992) 5657;\newline
 R.I. Nepomechie, J. Phys. A: Math. Gen. 33 (2000) L21.
\bibitem{open2}
 J.R. Links and M.D. Gould, Int. J. Mod. Phys. B 10 (1996) 3461;\newline
 A.J. Bracken, X.Y. Ge, Y.Z. Zhang and H.Q. Zhou,
   Nucl. Phys. B 516 (1998) 588;\newline
 H.Q. Zhou, Phys. Rev. B 53 (1996) 5089; Phys. Lett. A 228 (1997) 48.
\bibitem{open3}
 H. Fan, B.Y. Hou, K.J. Shi and Z.X. Yang,
   Nucl. Phys. B 478 (1996) 723;\newline
 M. Shiroishi and M. Wadati, J. Phys. Soc. Jpn. 66 (1997) 2288;\newline
 A. Foerster and M. Karowski, Nucl. Phys. B 408 (1993) 512;\newline
H.J. de Vega and A. Gonzalez Ruiz, J. Phys. A: Math. Gen. 26 (1993) L519.
\bibitem{open4}
H. Asakawa and M. Suzuki, Physica A 236 (1997) 376;
H. Asakawa,  Physica A 256 (1998) 229.
\bibitem{bqft}
 S. Ghoshal and A. Zamolodchikov, Int. J. Mod. Phys. A9 (1994) 3841.
\bibitem{q-w}
 A. Yacoby, R. Schuster and M. Meiblum, Phys. Rev. B 53 (1996) 9583;\newline
 V. T. Petrashov, V. N. Antonov, P. Delsing and T. Claeson,
 Phys. Rev. Lett. 74 (1995) 5268
\bibitem{imp2}
 Y. Wang, J. Dai, Z.-N. Hu and F.-C. Pu,
 Phys. Rev. Lett. 79 (1997) 1901;\newline
 Z.-N Hu  and F.-C. Pu, Phys. Rev. B 58 (1998) 2925.
\bibitem{imp3}
 H. Frahm and A.A. Zvyagin, J. Phys. Condens. Matter 9 (1997) 9939;\newline
 A.A. Zvyagin and H. Johannesson, Phys. Rev. Lett. 81 (1998) 2751;\newline
 G. Bed\"urftig and H. Frahm, J. Phys. A: Math. Gen. 32 (1999) 4585; \newline
   Physica E 4 (1999) 246;\newline
J. Links and A. Foerster,  J. Phys. A: Math. Gen. 32 (1999) 147;\newline
 X.-W. Guan, U. Grimm, R.A. R\"{o}mer and M. Schreiber,
   J. Phys. A: Math. Gen. 33 (2000) 3863.
\bibitem{imp4}
 H.Q. Zhou, X.-Y. Ge, J.R. Links and M.D. Gould,
   Nucl. Phys. B 546 (1999) 779;\newline
 H.-Q. Zhou and M.D. Gould, Phys. Lett. A 251 (1999) 279;\newline
 H.-Q. Zhou, X.-Y. Ge and M.D. Gould, J. Phys. A: Math. Gen. 32 (1999) 137.
\bibitem{Zv}
 B. Brendel, H. Frahm and R.M. Noack, Phys. Rev. B 58 (1998) 10225.
\bibitem{supal}
 A.J. Bracken, G.W. Delius, M.D. Gould and Y.Z. Zhang,
   J. Phys. A: Math. Gen. 27 (1994) 6551;\newline
 Z. Maassarani, J. Phys. A: Math. Gen. 28 (1995) 1305;\newline
 M.D. Gould, J.R. Links, Y.-Z. Zhang and I. Tsonhantjis,
   J. Phys. A: Math. Gen. 30 (1997) 4313.
\bibitem{deg}
 T. Deguchi, A. Fujii and K. Ito, Phys. Lett. B 238 (1990) 242.
\bibitem{doub}
 M.J. Martins and P.B. Ramos, Phys. Rev. B 56 (1997) 6376.
\bibitem{sces}
 H. Saleur, J. Phys. A: Math. Gen. 24 (1991) 1137;\newline
 A. Foerster and M. Karowski, Phys. Rev. B 46 (1992) 9234; \newline
 Nucl. Phys. B 396 (1993) 611.
\bibitem{Essler}
 F.H.L. Essler, V.E. Korepin and K. Schoutens, Phys. Rev. Lett.
   68 (1992) 2960; Phys. Rev. Lett. 70 (1993) 73.
\bibitem{LW}
 E.H. Lieb and F.Y. Wu, Phys. Rev. Lett. 20 (1968) 1445.
\bibitem{doubS}
 H. Saleur, J. Phys. A: Math. Gen. 32 (1999) L207.
\bibitem{wire}
 F. Lesage, H. Saleur and P. Simonetti, Phys. Rev. B 37 (1998) 4694.
\bibitem{MG1}
 M.J. Martins and X.-W. Guan, Nucl. Phys. B 562 (1999) 433.
\bibitem{thermo}
 A. Kl\"{u}mper and R.Z. Bariev, Nucl. Phys. B 458 (1995) 625;\newline
 G. J\"{u}ttner, A. Kl\"{u}mper and J. Suzuki,
   Nucl. Phys. B 487 (1997) 656;\newline
 C. Destri and H.J. de Vega, Phys. Rev. Lett. 69 (1992) 2313;
   Nucl. Phys. B 438 (1995) 413;\newline
 X. Zotos, P. Naet and P. Prelov, Phys. Rev. B 55 (1997) 11029;\newline
 M. Distasio and X. Zotos, Phys. Rev. Lett. 74 (1995) 2025.
\bibitem{Mar}
 M.J. Martins and  P.B. Ramos, Nucl. Phys. B 522 (1998) 413.
\bibitem{G1}
 X.-W. Guan, J. Phys. A: Math. Gen. 33 (2000) 5391;\newline
 A. Foerster, X.-W. Guan, J. Links, I. Roditi and H.-Q. Zhou,
 Nucl. Phys. B 596 (2001) 525.
\bibitem{J-W}E. Olmedilla, M. Wadati and Y. Akutsu,
 J. Phys. Soc. Jpn.  36 (1987) 2298.
\end{thebibliography}

\end{document}